\documentclass[letterpaper,twocolumn,prl,aps,superscriptaddress,amsmath,amssymb,floatfix]{revtex4-1}
\usepackage{mathptmx}
\usepackage{color}
\usepackage{verbatim}
\usepackage{float}
\usepackage{amsmath}
\usepackage{amssymb}
\usepackage{graphicx}
\usepackage{esint}
\usepackage[unicode=true,
 bookmarks=true,bookmarksnumbered=false,bookmarksopen=false,
 breaklinks=false,pdfborder={0 0 1},backref=false,colorlinks=true]
 {hyperref}
\hypersetup{
 linkcolor=magenta,urlcolor=blue,citecolor=blue,pdfstartview={FitH},hyperfootnotes=false}

\makeatletter

\usepackage{textcomp}
\usepackage{epstopdf}

\pdfpageheight\paperheight
\pdfpagewidth\paperwidth

\@ifundefined{textcolor}{}{%
 \definecolor{BLACK}{gray}{0}
 \definecolor{WHITE}{gray}{1}
 \definecolor{RED}{rgb}{1,0,0}
 \definecolor{GREEN}{rgb}{0,1,0}
 \definecolor{BLUE}{rgb}{0,0,1}
 \definecolor{CYAN}{cmyk}{1,0,0,0}
 \definecolor{MAGENTA}{cmyk}{0,1,0,0}
 \definecolor{YELLOW}{cmyk}{0,0,1,0}
}

\usepackage{xcolor}
\usepackage{soul}
\definecolor{blue}{rgb}{0,0,1}
\definecolor{red}{rgb}{1,0,0}
\definecolor{green}{rgb}{0,1,0}

\@ifundefined{textcolor}{}{%
 \definecolor{BLACK}{gray}{0}
 \definecolor{WHITE}{gray}{1}
 \definecolor{RED}{rgb}{1,0,0}
 \definecolor{GREEN}{rgb}{0,1,0}
 \definecolor{BLUE}{rgb}{0,0,1}
 \definecolor{CYAN}{cmyk}{1,0,0,0}
 \definecolor{MAGENTA}{cmyk}{0,1,0,0}
 \definecolor{YELLOW}{cmyk}{0,0,1,0}
}

\usepackage{xcolor}\usepackage{soul}
\definecolor{blue}{rgb}{0,0,1}
\definecolor{red}{rgb}{1,0,0}
\definecolor{green}{rgb}{0,1,0}

\usepackage{soul}

\usepackage{graphicx}
\usepackage{siunitx}

\makeatother

\begin{document}


\title{Enhancing single-atom loading in tightly confined dipole traps \\ with ancillary dipole beam}

\author{Guang-Jie~Chen}
\affiliation{Key Laboratory of Quantum Information, CAS, University of Science and Technology of China, Hefei, Anhui 230026, China\\}
\affiliation{CAS Center For Excellence in Quantum Information and Quantum Physics, University of Science and Technology of China, Hefei, Anhui 230026, China.}

\author{Zhu-Bo Wang}
\affiliation{Key Laboratory of Quantum Information, CAS, University of Science and Technology of China, Hefei, Anhui 230026, China\\}
\affiliation{CAS Center For Excellence in Quantum Information and Quantum Physics, University of Science and Technology of China, Hefei, Anhui 230026, China.}

\author{Chenyue Gu}
\affiliation{Key Laboratory of Quantum Information, CAS, University of Science and Technology of China, Hefei, Anhui 230026, China\\}
\affiliation{CAS Center For Excellence in Quantum Information and Quantum Physics, University of Science and Technology of China, Hefei, Anhui 230026, China.}

\author{Dong Zhao}
\affiliation{Department of optics and optics engineering, University of Science and Technology of China, Hefei, Anhui 230026, P. R. China\\}

\author{Ji-Zhe Zhang}
\affiliation{Key Laboratory of Quantum Information, CAS, University of Science and Technology of China, Hefei, Anhui 230026, China\\}
\affiliation{CAS Center For Excellence in Quantum Information and Quantum Physics, University of Science and Technology of China, Hefei, Anhui 230026, China.}

\author{Yan-Lei Zhang}
\affiliation{Key Laboratory of Quantum Information, CAS, University of Science and Technology of China, Hefei, Anhui 230026, China\\}
\affiliation{CAS Center For Excellence in Quantum Information and Quantum Physics, University of Science and Technology of China, Hefei, Anhui 230026, China.}

\author{Chun-Hua Dong}
\affiliation{Key Laboratory of Quantum Information, CAS, University of Science and Technology of China, Hefei, Anhui 230026, China\\}
\affiliation{CAS Center For Excellence in Quantum Information and Quantum Physics, University of Science and Technology of China, Hefei, Anhui 230026, China.}

\author{Kun Huang}
\email{huangk17@ustc.edu.cn}
\affiliation{Department of optics and optics engineering, University of Science and Technology of China, Hefei, Anhui 230026, P. R. China\\}

\author{Guang-Can Guo}
\affiliation{Key Laboratory of Quantum Information, CAS, University of Science and Technology of China, Hefei, Anhui 230026, China\\}
\affiliation{CAS Center For Excellence in Quantum Information and Quantum Physics, University of Science and Technology of China, Hefei, Anhui 230026, China.}

\author{Chang-Ling Zou}
\email{clzou321@ustc.edu.cn}
\affiliation{Key Laboratory of Quantum Information, CAS, University of Science and Technology of China, Hefei, Anhui 230026, China\\}
\affiliation{CAS Center For Excellence in Quantum Information and Quantum Physics, University of Science and Technology of China, Hefei, Anhui 230026, China.}




\date{\today}

\begin{abstract}
Single atoms trapped in tightly focused optical dipole traps provide an excellent experimental platform for quantum computing, precision measurement, and fundamental physics research. In this work, we propose and demonstrate a novel approach to enhancing the loading of single atoms by introducing a weak ancillary dipole beam. The loading rate of single atoms in a dipole trap can be significantly improved by only a few tens of microwatts of counter-propagating beam. It was also demonstrated that multiple atoms could be loaded with the assistance of a counter-propagating beam. By reducing the power requirements for trapping single atoms and enabling the trapping of multiple atoms, our method facilitates the extension of single-atom arrays and the investigation of collective light-atom interactions.
\end{abstract}

\maketitle


\section{\label{sec:level1}Introduction}

In recent decades, trapping single neutral atoms has attracted significant attention for its potential applications in studies of matter waves~\cite{Kaufman2014,Brown2023}, light-atom interactions~\cite{Tey2008,OShea2013,Chin2017,Deist2022,Liu2023,Will2021,Zhou2023}, and quantum information sciences~\cite{Saffman2019,Browaeys2020,Morgado2021,Wu2021}. One of the most important approaches for obtaining single neutral atoms is via a tight dipole trap~\cite{Grimm2000,Kuppens2000}, the so-called single atom tweezer, which is generated by focusing red detuned dipole light into a several-micron waist beam spot through a high numerical aperture ($\mathrm{N.A.}$) objective~\cite{Haokang2020}. In such a small dipole trap, the light-assisted collision blockade effect leaves the trap with only 0 atoms and 1 atom, allowing probabilistic preparation of a single atom~\cite{Schlosser2001,Schlosser2002,Fung2015}. Recently, a vital breakthrough was achieved in this research field, as defect-free single-atom arrays could be prepared in a nearly deterministic manner through the rearrangement of single-atom tweezers~\cite{Barredo2016,Endres2016}. In particular, researchers have used two-dimensional acousto-optic deflectors and spatial light modulators to scale the number of dipole traps to tens or even hundreds~\cite{Bernien2017,Kaufman2021,Ebadi2021}, and can arbitrarily rearrange the dipole traps according to experimental needs~\cite{Sheng2022,Barredo2018,Labuhn2016,Omran2019}. Deterministic defect-free single-atom arrays provide a promising platform for quantum simulation~\cite{Bernien2017,Ebadi2021} and quantum computing~\cite{Shi2022,Bluvstein2022,Graham2022}. 

Alternatively, neutral atom arrays with optical lattices have been extensively investigated~\cite{Bloch2008,Gross2017,Yang2020a,Yang2020}. One- or two-dimensional optical lattices, as arrays of optical dipole trap potentials with each site, can probabilistically load single atoms, and can be prepared by the interference of two or more dipole trap beam that propagates along different directions~\cite{Piotrowicz2013}. Such an approach allows capturing a larger number of atoms more easily than dipole trap arrays, which is beneficial for applications where a large number of atoms are needed, such as quantum simulations of many-body physics and optical lattice atomic clocks~\cite{Campbell2017,Nemitz2016,Takano2016}. 
Compared to an optical tweezer, the essential difference of a 1D optical lattice is a counter-propagating dipole beam is applied. Previously, it was found that partially reflected dipole beam could effectively modify the atom loading dynamics in a dipole trap~\cite{Wang2023}. It is curious whether the crossover from the single-atom trapping to optical lattices could be observed if the power of the counter propagating dipole trap beam applied to an optical dipole trap is adjusted.

In this work, we experimentally investigated the impact of ancillary dipole beam on atom loading in a tightly confined optical dipole trap, and demonstrated the crossover from the ordinary dipole trap for single atoms trapping to the 1D lattice for multiple atom loading, by varying the power of the ancillary dipole beam ($P_{\mathrm{anc}}$). We found that the crossover can be divided into two regimes, the low-power regime and the high-power regime. In the low-power regime, the single atom loading rate and lifetime are very sensitive to $P_{\mathrm{anc}}$. Only dozens of microwatts can substantially increase the single-atom loading rate from near $0\%$ to near $50\%$ and improve the lifetime from nearly \SI{100}{ms} to nearly \SI{1000}{ms}. In the high-power regime, as $P_{\mathrm{anc}}$ increases, the normal dipole trap gradually transforms into a standing-wave dipole trap that can capture 2 or more atoms. This approach provides a way to prepare a few-atoms ensemble that may be beneficial for research on the collective effect of atoms~\cite{Vochezer2018} and might also lead to new ideas for increasing the loading probabilities of single atoms in tweezer arrays.

\begin{figure*}[t]
\centering
\includegraphics[width=0.7\linewidth]{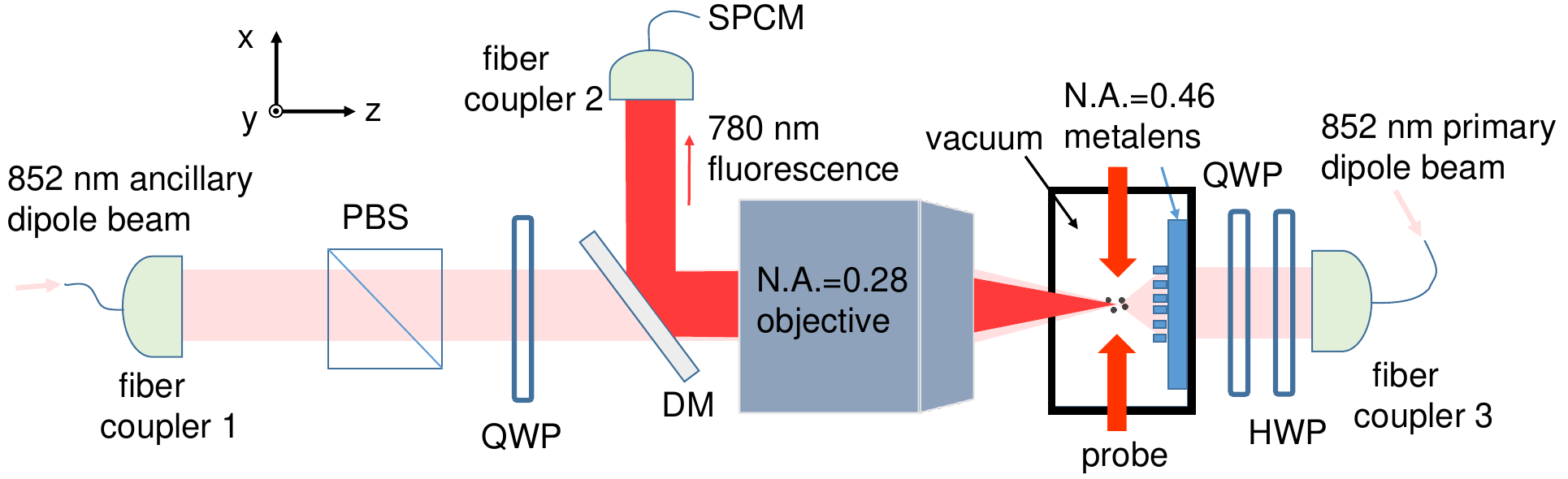}
\caption{The experimental setup for investigating the single atom loading dynamics. The dipole beam is provided by a \SI{852}{nm} laser. The  dipole beam output from fiber coupler 3 is focused by a metalens ($\mathrm{N.A.}=0.46$), with its polarization is controlled by a half-wave plates (HWP) and a quarter-wave plates (QWP). The ancillary dipole beam output from fiber coupler 1 is focused by a commercial objective ($\mathrm{N.A.}=0.28$), with its polarization is controlled by a polarization beam splitter (PBS) and a QWP. All the fibers in this experiment are polarization maintaining fibers, and the beam's polarization have been matched to the principal axis of the fibers. The waists of two Gaussian dipole laser beams overlap in the magneto-optical trap (MOT). The fluorescence from the trapped atoms in the dipole trap is collected by the objective and detected by a single photon count module (SPCM) through a dichroic mirror (DM) and fiber coupler 2.
}
\label{fig1}
\end{figure*}

\section{\label{sec:level2}Experimental Setup}

Our experimental setup is depicted in Fig.~\ref{fig1}. We utilized a tightly confined optical dipole trap for single-atom optical tweezers, featuring a waist of approximately $1.3\,\mathrm{\mu m}$. This trap is created by a high numerical aperture($\mathrm{N.A.}=0.46$) metalens~\cite{ChenGJ2023,Hsu2022} with a working distance of $2.6\,\mathrm{mm}$. The metalens, composed of a thin layer of silicon nanostructures (500\,nm) on the surface of 500\,$\mu$m sapphire, is directly placed in a glass vacuum cell. In particular, our metalens not merely serves as a replacement for a conventional objective lens, but is also designed to be multifunctional and be able to polarize the transmitted beam. It consistently outputs left-handed circularly polarized beam that is insensitive to the polarization of the input beam. 

To prepare cold $^{87}$Rb atom clouds with a temperature of about $100\,\mu\mathrm{K}$, a conventional magneto-optic trap (MOT, not shown in Fig.~\ref{fig1}) is employed. The atom cloud is centered around $2.5\,\mathrm{mm}$ away from the nanostructure on the chip to overlap with the focus of the metalens. As shown in Fig.~\ref{fig1}, optical dipole beam, provided by an $852\,\mathrm{nm}$ laser, is coupled to the metalens through fiber coupler 3 to produce the primary dipole trap. The polarization of the input beam to the metalens is tuned by a quarter-wave plate (QWP) and a half-wave plate (HWP). 
An additional polarization-gradient cooling (PGC) procedure was applied in our experiments to further cool the atomic temperature to around $50\,\mu\mathrm{K}$~\cite{Schlosser2002,ChenGJ2023}.

The trapped atoms are illuminated by a probe beam and its retroreflective beam, which are nearly parallel to the metalens chip. The probe beam comprises two laser frequencies, which are the same as the cooling and repump beam of the MOT, with the powers are about $220\,\mu\mathrm{W}$ and $80\,\mu\mathrm{W}$, respectively. The probe beam is linearly polarized along the $y$-direction [Fig.~\ref{fig1}]. In order to counteract the potential impact of a fictitious magnetic field induced by the circularly polarized dipole trap, a bias magnetic field of $0.7\,\mathrm{G}$ is applied in the $z$ direction. We used a commercial objective lens, $\mathrm{N.A.}=0.28$ (PDV M Plan APO L), to collect the fluorescence of single atoms at $780\,\mathrm{nm}$. The fluorescence signal is then directed to fiber coupler 2 and measured utilizing a single-photon count module (SPCM, Excelitas SPCM-AQRH-14-FC) with an integration time of \SI{50}{ms}. 
The objective lens is achromatic at both the $780\,\mathrm{nm}$ and $852\,\mathrm{nm}$ wavelengths, allowing for independent separation, adjustment, and characterization of the dipole beam from the fluorescence measurement by introducing a dichroic mirror (DM, Semrock) into the optical setup. As illustrated in the left part of Fig.~\ref{fig1}, the primary dipole beam of the metalens can be collected through an optical collimation setup. The polarization of the primary dipole trap was also verified by measuring the collected dipole beam through the objective of a polarization analyzer (Schafter Kirchhoff, SK010PA). We found that no matter how the polarization of the incident beam changes, the polarization of the focused beam is very close to that of left-handed circularly polarized beam, and the ellipticity $\eta$ of the focused beam is between 0.6 and 0.7. 
It noteworthy that all the fibers ultilized in our experiments are polarization-maintaining fibers, and their principle axes are aligned to match the direction of linearly polarized beam, as determined by the polarization analyzer.

\begin{figure*}[!t]
\centering
\includegraphics[width=0.8\linewidth]{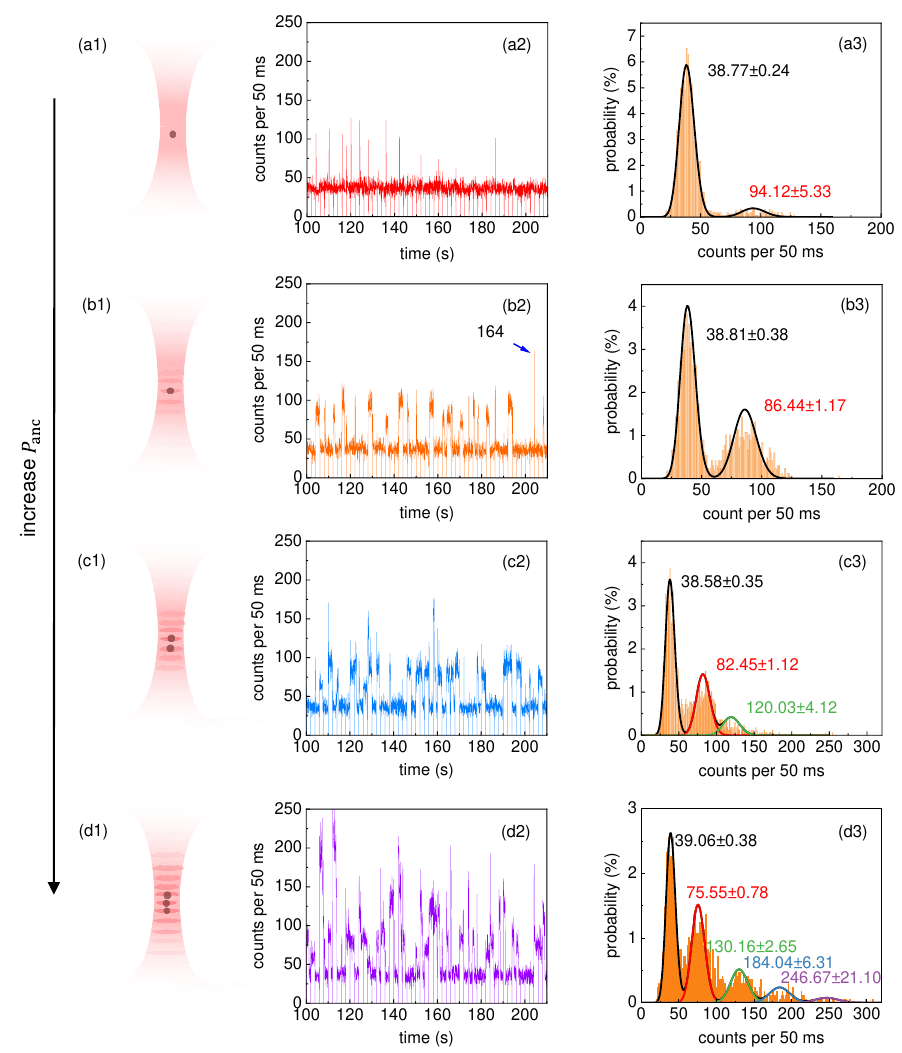}
\caption{Illustrations of dipole traps with varying ancillary dipole beam power $P_{\mathrm{anc}}$, and the corresponding experimental results of atomic fluorescence signals and histogram statistics of counts.  Panels (a1)-(a3), (b1)-(b3), (c1)-(c3), and (d1)-(d3) show the results for $P_{\mathrm{anc}}=11\,\mathrm{{\mu}W}$,  $P_{\mathrm{anc}}=92\,\mathrm{{\mu}W}$, $P_{\mathrm{anc}}=238\,\mathrm{{\mu}W}$, and  $P_{\mathrm{anc}}=476\,\mathrm{{\mu}W}$, respectively, while the primary dipole beam power is fixed at $P_{\mathrm{p}}=10\,\mathrm{mW}$. In panels (a3)-(d3), the lines represent Poisson distribution fitting of photon counts, and the labels are the corresponding fitted mean counts of 
trapped atoms.}
\label{fig2}
\end{figure*}

To optimize the overlap between the focused ancillary beam from the objective lens and the primary dipole beam from the metalens, we collect the primary dipole trap beam from the metalens into fiber coupler 1. The measured coupling efficiency allows us to achieve maximal overlap between the two beams, in accordance with the time-reversal symmetry of dipole beam propagation in our setup. This maximal overlap results in the ancillary dipole beam exerting the strongest effect on the dipole trap. As the two lenses are respectively fixed on different holders, environmental temperature fluctuations and vibrations could induce misalignment between the focuses, thereby affecting the strength of the collected signal. To mitigate these effects, we monitor the primary dipole beam transmittance and retain only data with a transmittance above a certain threshold ($60\%$ of the maximum), thereby achieving passive stabilization of the coupling. 

In our experiments, the experimental procedure was repeated many times, with each cycle lasting 6 seconds. During the first 3 seconds, the MOT and dipole beam are turned on to concentrate atoms from the environment and load single atoms into the dipole trap. 
Subsequently, the MOT is switched off, and the atoms trapped in the dipole trap are characterized by collecting and counting the fluorescence signal for a period of 2 seconds. Finally, all lasers are switched off with an additional 1 second allocated for redundancy and for emptying the dipole trap. We set the first data point of the fluorescence signal to 0 as the starting point of each cycle. Similar to previous studies~\cite{Wang2023,Schlosser2001}, the fluorescence signal exhibits telegraph chart-like traces, with sudden rises and drops of the signal indicating the loading and escaping events of single atoms in the dipole trap. We perform various experiments under different experimental conditions by varying the power of the primary dipole beam focused by the metalens, denoted as $P_{\mathrm{p}}$, and the power of the ancillary dipole beam, denoted as $P_{\mathrm{anc}}$. Both powers are calibrated to the powers of the focused beam in vacuum.

\section{\label{sec:level3}Results}

The typical fluorescence signal of single atoms in the optical dipole trap was experimentally investigated with various ancillary dipole beam powers $P_{\mathrm{anc}}$, as schematically illustrated in Fig.~\ref{fig2}(a2)-(d2). Panels (a2)-(d2) display segments of the collected photon counting signal trace for different $P_{\mathrm{anc}}$, while the power of the primary dipole trap beam $P_{\mathrm{p}}$ by the metalens is fixed at $10\,\mathrm{mW}$. As $P_{\mathrm{anc}}$ increases, both the number and height of steps in the fluorescence in the  dipole trap change significantly. The analysis of the histogram statistics of the entire fluorescence signal is presented in Fig.~\ref{fig2}(a3)-(d3). These histograms were fitted using the summation of Poisson distributions $\sum _{i=0}^n {C_i} e^{-\lambda_i }\lambda_i ^k/k!$~\cite{Schlosser2001}, where each Poisson distribution with a fitted $\lambda_i$ corresponds to the fluorescence signal of different numbers of trapped atoms. Here, $C_i$ indicates the probability of trapping $i$ atoms, satisfying the normalization condition $\sum _{i=0}^n {C_i}=1$ and $k$ indicates different counts of the fluorescence signal. The difference between the mean value of each Poisson distribution $\lambda_i$ represents the collected fluorescence of a single atom within $50\,\mathrm{ms}$.

In Fig.~\ref{fig2}, panels (a1)-(a3) display the results for $P_{\mathrm{anc}}=11\,\mathrm{{\mu}W}$. The probability of trapping a single atom is ${8.2\pm 3.2\%}$, and the lifetime of a single atom, i.e. the mean width of the steps in the fluorescence traces, is only about 50-100\,ms. As $P_{\mathrm{anc}}$ increases to $92\,\mathrm{{\mu}W}$ in panels (b1)-(b3), the probability of trapping a single atom increases to ${30.3 \pm 3.7\%}$, and the steps last longer with an averaging duration of about 300\,ms. Compared to panels (a1)-(a3), the trapping probability and lifetime of a single atom in the dipole trap are significantly enhanced even when $P_{\mathrm{anc}}$ is less than $1\%$ of $P_{\mathrm{p}}$. Additionally, a notably high photon count signal of 164 per $\SI{50}{\milli\second}$ is observed in panel (b2), which is much greater than the mean fluorescence count of a single atom. This suggests that two or more atoms have been simultaneously trapped in the dipole trap for a very short duration. Nonetheless, in most cases, the dipole trap can still only accommodate either 0 atoms and 1 atom. Panels (c1)-(c3) and (d1)-(d3) show the results for further increases in $P_{\mathrm{anc}}$. The occurrence of counts exceeding the mean count of a single atom becomes more frequent. In panels (c2) and (d2), sustained high count steps for multiple atoms are observed, indicating suppression of the collision blockade effect, which typically results in the simultaneous loss of two atoms~\cite{Schlosser2001,Schlosser2002}. Therefore, the trapped atoms in the dipole trap do not immediately collide with each other, allowing the dipole trap to accommodate multiple atoms in a short duration. Compared to previous studies on modified two-atom collision rate in dipole traps due to the retro-reflection of dipole beam~\cite{Wang2023}, our results reveal more significant two-, three-, and even more atom steps. This indicates a gradual change in the dipole trap from a conventional single-atom tweezer to a standing-wave optical lattice, where each anti-node position can potentially serve as a secondary dipole trap to capture a single atom, as depicted in panels (c1)-(d1).

\begin{figure}[!t]
\centering
\includegraphics[width=1\linewidth]{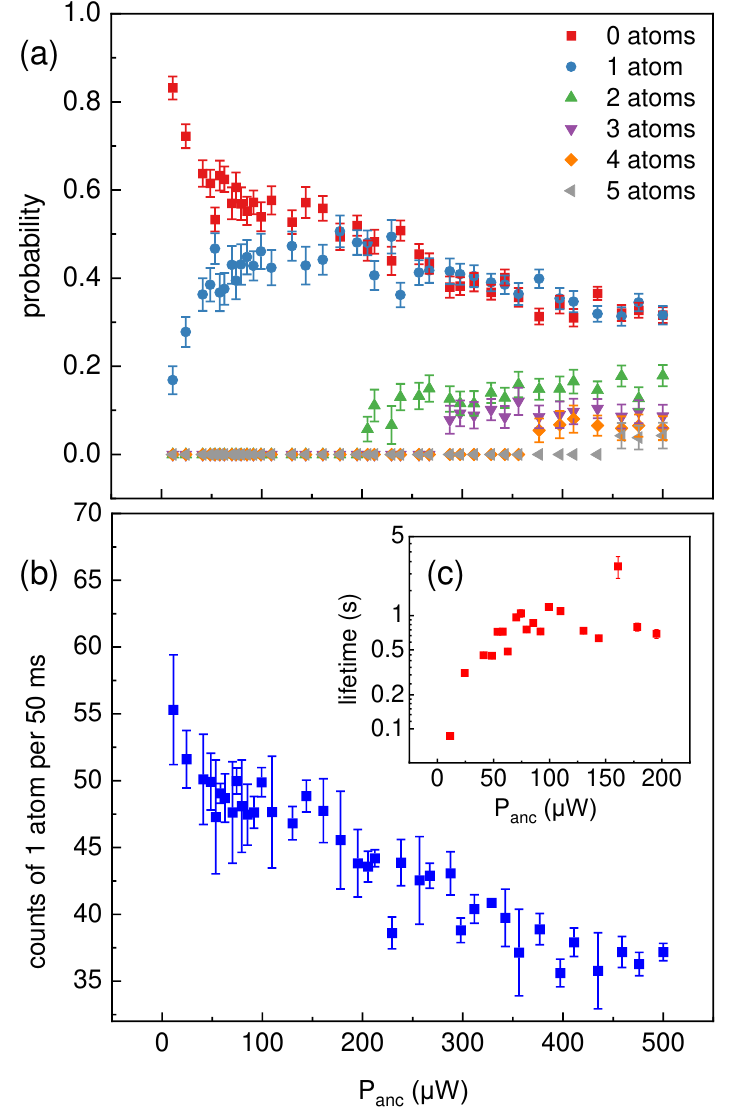}
\caption{The experimental results for trapping single atoms in a dipole trap affected by different ancillary dipole beam powers $P_{\mathrm{anc}}$. (a) The probabilities of trapping $n$ atoms in the dipole trap affected by $P_{\mathrm{anc}}$, with $n=0,1,2...$. (b) The fluorescence signal step of a single trapped atom against $P_{\mathrm{anc}}$. (c) The lifetime of a trapped single atom varying with $P_{\mathrm{anc}}$ when $P_{\mathrm{anc}}<200\mathrm{{\mu}W}$ in the low-power regime. }
\label{fig3}
\end{figure}



To systematically investigate the influence of the power of ancillary dipole beam $P_{\mathrm{anc}}$ on the performance of trapping atoms in a dipole trap, we varied $P_{\mathrm{anc}}$ every \SI{6}{s} using a variable metallic neutral density filter. We collected the fluorescence signal cyclically to minimize the effects of  parameter drift from the environment. Figure~\ref{fig3} presents the statistical data on the probability of trapping atoms, the lifetime, and the fluorescence counts of trapped single atoms when $0\ \mathrm{{\mu}W}<P_{\mathrm{anc}}<500\,\mathrm{{\mu}W}$. According to Fig.~\ref{fig3}(a), the dipole trap can only trap single atoms when $P_{\mathrm{anc}}<200\,\mathrm{{\mu}W}$, and we refer to this as the low-power regime. In this regime, the collision blockade effect ensures that only a single atom can be trapped in a dipole trap. The atom loading probability increases rapidly and saturates as $P_{\mathrm{anc}}$ increases, similar to the effect of increasing $P_{\mathrm{p}}$ in ordinary dipole traps, but with a more pronounced change. A detailed comparison is provided in Fig.~\ref{fig4} and Fig.~\ref{fig5}. 

When further increasing the power of ancillary dipole beam to $P_{\mathrm{anc}}>200\,\mathrm{{\mu}W}$, the dipole trap can trap multiple atoms, and we call this area as the high-power regime. As shown in Fig.~\ref{fig3}(a), the probability of trapping different numbers of atoms emerges when $P_{\mathrm{anc}}$ exceeds a certain threshold and eventually saturates. We observed that below the threshold power of $P_{\mathrm{anc}}$ for $i$ atoms, the number of atoms in the dipole trap could not exceed $i$. For instance, in Fig.~\ref{fig3}(a), when $205\,\mathrm{{\mu}W}<P_{\mathrm{anc}}<267\,\mathrm{{\mu}W}$, the dipole trap can only capture up to 2 atoms most of the time, with an extremely low probability of collecting a higher signal. For example, there is a peak of 250 per $50\,\mathrm{ms})$ counts in Fig.~\ref{fig2}(c3). 

Figure~\ref{fig3}(b) shows the magnitude of fluorescence with $P_{\mathrm{anc}}$ for a single atom collected by SPCM within a duration of $50\,\mathrm{ms}$. As $P_{\mathrm{anc}}$ increases, the fluorescence gradually decreases, indicating that the AC-Stark shift or dipole trap depth felt by the atoms has been strongly modified even if the power of ancillary propagating dipole beam ($500\,\mathrm{{\mu}W}$) is only 1/20 of the primary dipole beam power ($10\,\mathrm{mW}$). 
We note that the fluctuations in the alignment of the two lenses might impact the collection of fluorescence signal photon counts, but only contribute to the variance of the signal. Figure~\ref{fig3}(c) shows the lifetime of the captured single atoms with $P_{\mathrm{anc}}$ in the low-power regime, while the lifetime of the multiple atoms case is difficult to calibrate. The lifetime increases drastically with increasing $P_{\mathrm{anc}}$, similar to the single-atom trapping probability, and saturates at $P_{\mathrm{anc}}\approx100\,\mathrm{{\mu}W}$.

\subsection{\label{sec:level4}The low-power regime}
\begin{figure}[!t]
\centering
\includegraphics[width=1\linewidth]{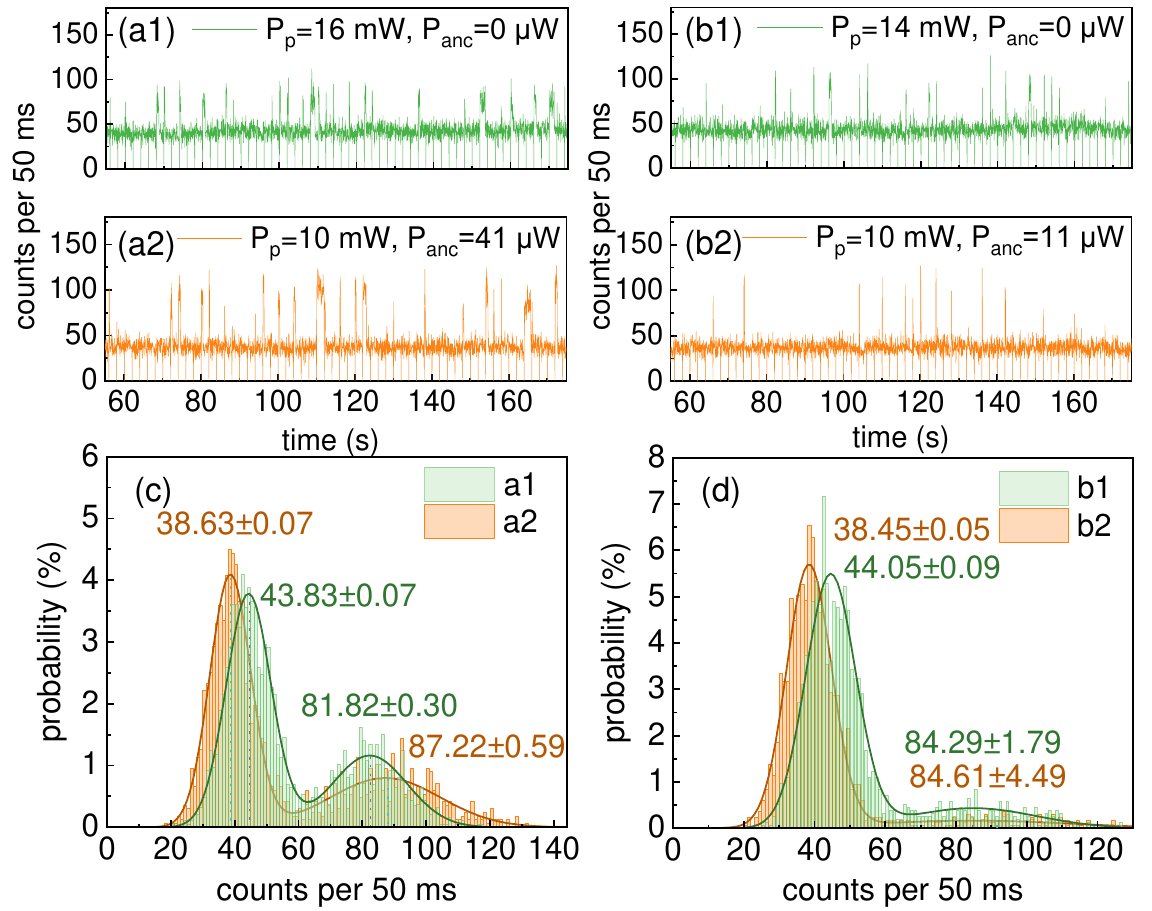}
\caption{The fluorescence signals obtained at different power levels of and ancillary dipole beams. (a1)(a2) show the signals for $P_{\mathrm{p}}=10\,\mathrm{mW}$,  $P_{\mathrm{anc}}=41\,\mathrm{{\mu}W}$ (orange) and $P_{\mathrm{p}}=10\,\mathrm{mW}$, $P_{\mathrm{anc}}=0\,\mathrm{{\mu}W}$ (green), while (b1) and (b2) show the signals for $P_{\mathrm{p}}=10\,\mathrm{mW}$, $P_{\mathrm{anc}}=11\,\mathrm{{\mu}W}$ (orange) and $P_{\mathrm{p}}=10\,\mathrm{mW}$, $P_{\mathrm{anc}}=0\,\mathrm{{\mu}W}$ (green). (c) and (d) The corresponding histogram statistics for the results presented in (a1)(a2) and (b1)(b2), respectively.}
\label{fig4}
\end{figure}

\begin{figure*}[!t]
\centering
\includegraphics[width=0.7\linewidth]{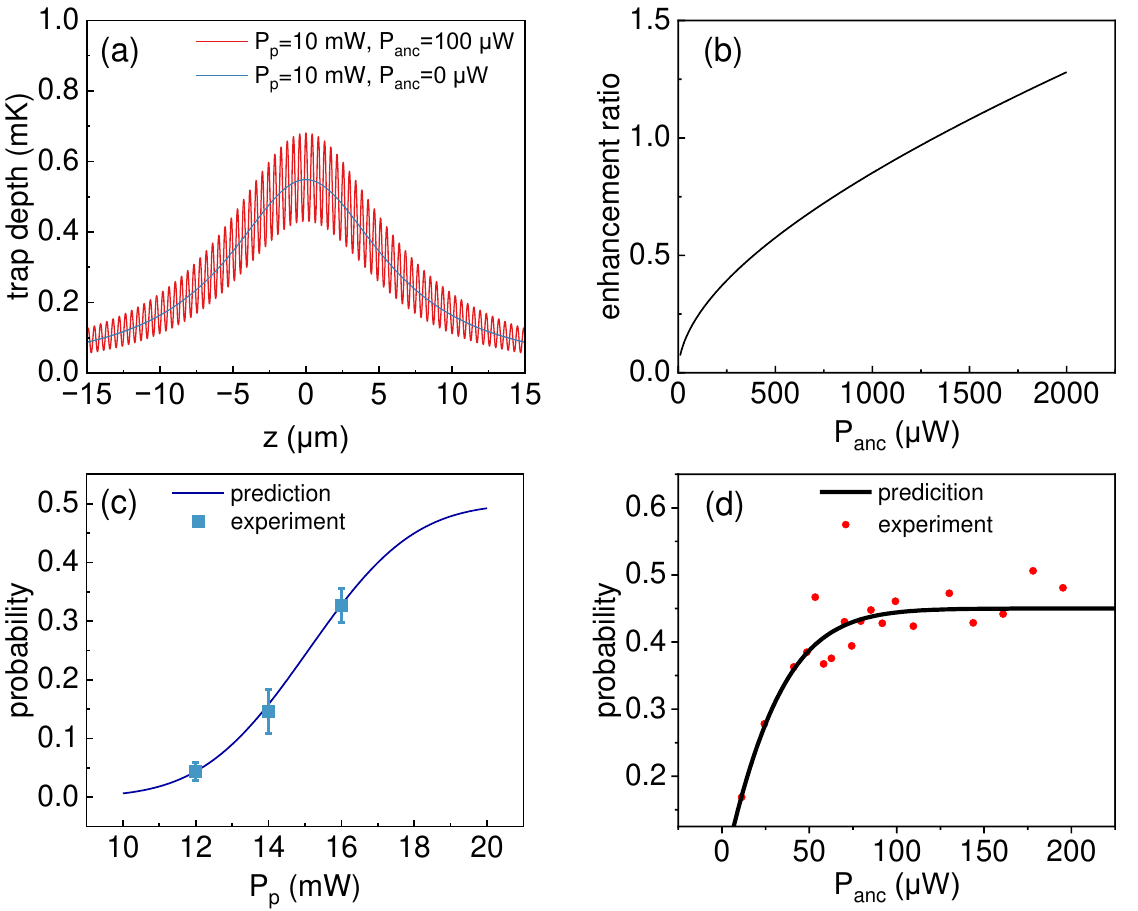}
\caption{The characterization of dipole trap. (a) Theoretical results of axial trap depth when dipole beam power is $10\,\mathrm{mW}$ and ancillary dipole beam power is $P_{\mathrm{anc}}=100\,\mathrm{{\mu}W}$ (red line) or $0$ (blue line).  (b) Calculated dipole trap depth enhancement ratio compared to the case without ancillary dipole beam. (c) Single-atom loading probability against $P_{\mathrm{p}}$ without ancillary dipole beam ($P_{\mathrm{anc}}=0$), with the dots from experiments and the line from a theoretical model. (d) The single-atoms loading probability against the power of ancillary dipole beam $P_{\mathrm{anc}}$, with $P_{\mathrm{p}}=10\,\mathrm{mW}$ is fixed. The dots are from experiments, and the black lines show the fitting with theoretical model.
}
\label{fig5}
\end{figure*}

The physics of the drastic dependence of atom loading probabilities on the ancillary dipole beam power ($P_{\mathrm{anc}}>0$) is further explored by comparing it with conventional traveling wave optical single-atom tweezers ($P_{\mathrm{anc}}=0$). As shown in Fig.~\ref{fig4}(a), the ability to load single atoms using an ancillary dipole beam with $P_{\mathrm{anc}}=41\,\mathrm{{\mu}W}$ for $P_{\mathrm{p}}=10\,\mathrm{mW}$ can be achieved by increasing the primary dipole beam power to $P_{\mathrm{p}}=16\,\mathrm{mW}$ when $P_{\mathrm{anc}}=0$, indicating that the loading probability and lifetime of single atoms are similar when either $41\,\mathrm{{\mu}W}$ of ancillary dipole beam or $6\,\mathrm{mW}$ of additional primary dipole beam is introduced to the system. 
Figure~\ref{fig4}(c) presents a histogram comparison of two sets of data in Fig.~\ref{fig4}(a). Both sets of data exhibit only two states, 0 atoms and 1 atom. In order to improve the R-squared value of the fit, we employed Gaussian distribution fitting~\cite{Kim}. In Fig.~\ref{fig4}(c), the two sets of data are comparable as the single atom loading probability for both sets of data is approximately 30\%, indicating a 37.5\% power reduction in the total dipole beam. This reduction in the power requirement for single-atom dipole trap may be beneficial for their extension to tweezer arrays~\cite{Barredo2016,Endres2016}. Additionally, due to the objective lens also collects dipole beam focused by the metalens and the 852\,nm dipole beam cannot be completely filtered by our interference filter, the fluorescence signal background increase by approximately 5.2 per $\SI{50}{\milli\second}$ for $P_{\mathrm{p}}=16\, \mathrm{mW}$ compared with $P_{\mathrm{p}}=10\,\mathrm{mW}$. 

Similarly, Fig.~\ref{fig4}(b) demonstrates that the performance of loading single atoms induced by increasing the power of the dipole trap by $4\,\mathrm{mW}$ could be realized by only introducing a $11\,\mathrm{{\mu}W}$ ancillary dipole beam. 
In Fig.~\ref{fig4}(d), a comparison of the histogram distributions of the fluorescence signal for the two cases of $P_\mathrm{p}=14\ \mathrm{mW}$ and $P_{\mathrm{anc}}=0\,\mathrm{{\mu}W}$ versus $P_{\mathrm{p}}=10\,\mathrm{mW}$ and $P_{\mathrm{anc}}=11\,\mathrm{{\mu}W}$ shows a power reduction of 28.5\% for achieving a single-atom loading probability of approximately 10\%. 

In both scenarios, we observed that the introduction of ancillary dipole beam led to a more pronounced single atoms fluorescence step when the system achieved a similar probability of trapping a single atom.  For instance, the fluorescence count step was measured at 48.59 per $\SI{50}{\milli\second}$ with ancillary dipole beam, while it is only  37.99 per $\SI{50}{\milli\second}$ for a conventional dipole trap, as shown in Fig.~\ref{fig4}(c). We speculate that the interference between the primary and ancillary dipole beam generated a standing-wave-like pattern, thereby enhancing the depth of the dipole trap and improving the performances in trapping a single atom. This resulted in an increase in the fluorescence collection efficiency due to the reduced range of atom motion along the beam axial direction, as the trap potential is modulated by the ancillary dipole beam compared to the normal traveling-wave dipole trap ($p_{\mathrm{anc}}=0$). 

\begin{figure*}[!t]
\centering
\includegraphics[width=1\linewidth]{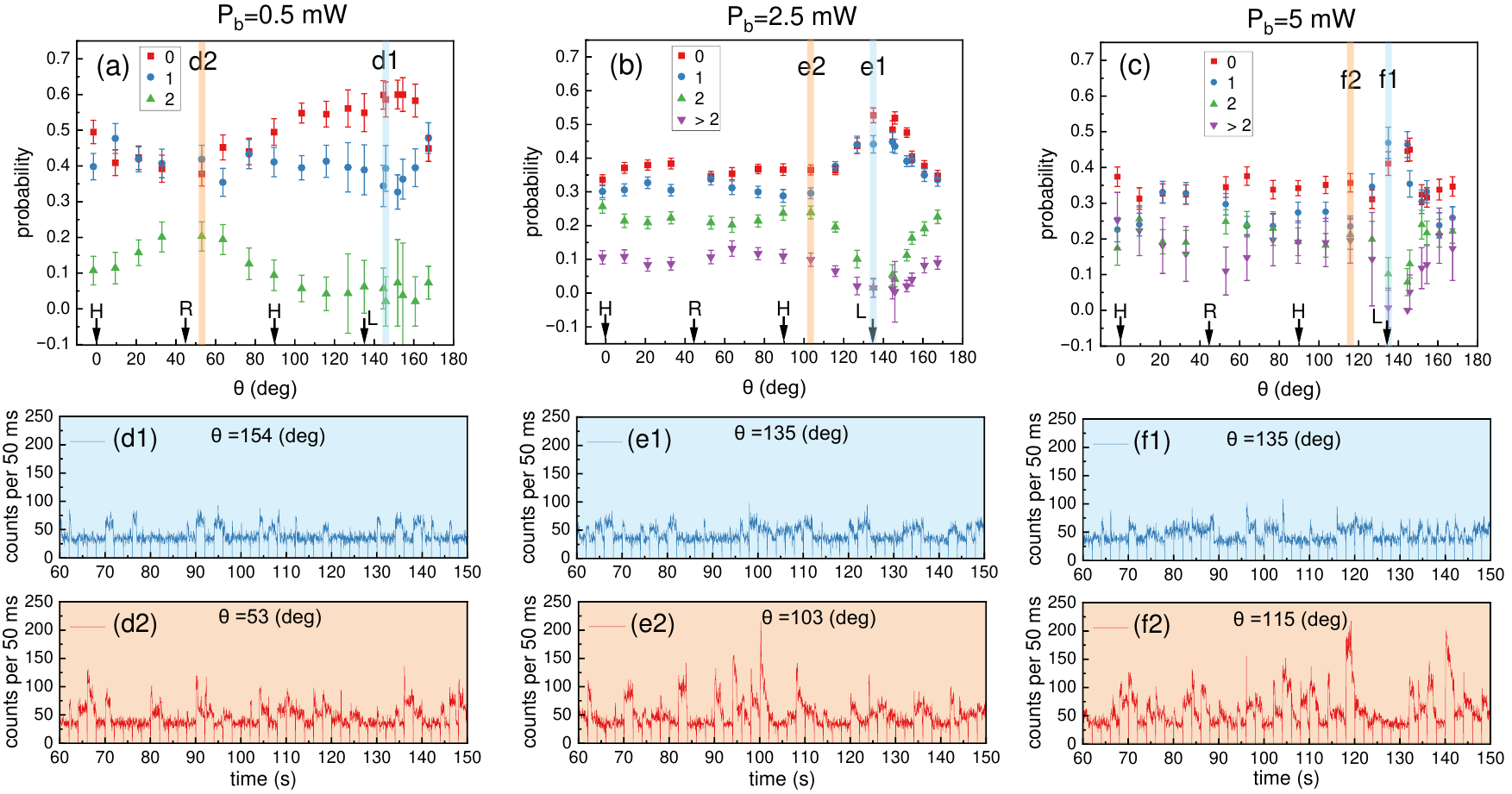}
\caption{The performance of dipole trap under a ancillary dipole beam with varying polarization. (a) and (d1)(d2),  (b) and (e1)(e2), and (c) and (f1)(f2) are the results for $P_{\mathrm{anc}}=0.5\,\mathrm{mW}$, $2.5\,\mathrm{mW}$ and $5\,\mathrm{mW}$, respectively, while the power of primary dipole beam is fixed at $10\,\mathrm{mW}$.
}
\label{fig6}
\end{figure*}

To quantitatively analyze the effect of the ancillary dipole beam, we theoretically calculate the modulation of the dipole trap potential. First, we assume that the electric field intensity distribution of the primary Gaussian beam along the axial direction ($z$-direction) is 
\begin{equation}
E(r,z)=a\frac{w_0}{w(z)} e^{\left[-\frac{r^2}{w(z)^2}-i \frac{2 \pi}{\lambda}  \left(\frac{r^2}{2R}+z\right)\right]}
\label{equ1}
\end{equation}
Here, $w(z)={w_0}\sqrt{1+\left(z/z_R\right)^2}$, $w_0$ is the beam waist, $z_R=\pi  w_0^2/\lambda$ is the Rayleigh length, $\lambda=852\,\mathrm{nm}$ is the dipole laser wavelength, $R=z\sqrt{1+\left(z_R/z\right)^2}$, and $a$ is a fitting coefficient that is determined by the dipole beam power in the experiments. The metalens is not a standard lens, we directly characterized the beam profiles around the focusing spots of the metalens. We approximately treat the primary dipole trap as a standard Gaussian profile with beam waist $w_{\mathrm{p}}=1.3\,\mu\mathrm{m}$ and $z_R=11.7\,\mu\mathrm{m}$, with the two parameters obtained by fitting the field distribution along the $r$-direction and $z$-direction, respectively. Such an equivalent Rayleigh length $z_R$ predicts a Gaussian beam waist of $w=1.78\,\mathrm{\mu m}$, which does not match the observed beam waist at the focus. This discrepancy is attributed to the deviation of the beam profile from the standard Gaussian distribution on the focal plane, and the deviations can be explained as the beam generated by metalens is effectively as a superposition of multiple Laguerre-Gaussian beams with different orders. Their combined effect makes the overall Rayleigh length appear to be elongated. It is also for this reason that the energy of the beam we use to capture single atoms only accounts for a part of the incident beam. Since the atoms are confined mainly around the maximum of the laser intensity distributions ($r=0$). Assuming that the effective portion of the laser power in the local tightly confined dipole trap beam is $\zeta$, the trap depth $U=\rho_{\mathrm{p}}/\rho_0\,\mathrm{mK}$ can be derived, where $\rho_0=2.17\,\mathrm{mW/\mu m^2}$ and the center dipole laser power density $\rho_{\mathrm{p}}=2\zeta P_{\mathrm{p}}/\pi w_{\mathrm{p}}^2$. According to our measurement of the AC Stark shift of about $20.7\,\mathrm{MHz}$ for the D2 transition of atoms in the trap when $P_{\mathrm{p}}=16.7\,\mathrm{mW}$, we estimate  $\zeta=0.33$  and plot the corresponding trap depth along the axial direction in Fig.~\ref{fig5}(a). 

When introducing ancillary dipole beam, the maximum optical power density becomes $\left( \sqrt{\rho_\mathrm{anc}}+\sqrt{\rho_{\mathrm{p}}} \right)^2$, where $\rho_\mathrm{anc}=2 P_{\mathrm{anc}}/\pi w_{\mathrm{anc}}^2$ due to the ancillary light and $w_{\mathrm{anc}}=2.03\,\mathrm{\mu m}$ for the objective lens. Here, we assume that the primary and ancillary dipole beam are aligned perfectly. The modulated optical potential is calculated and plotted in Fig.~\ref{fig5}(a), with $P_{\mathrm{p}}=10\,\mathrm{mW}$ and $P_{\mathrm{anc}}=100\,\mathrm{{\mu}W}$. A significantly enhancement of the trap depth by a factor of $13\%$ is observed for $P_{\mathrm{anc}}=100\,\mathrm{{\mu}W}$, compared to that for the case without it. The enhancement ratio by ancillary dipole beam can be directly approximated as  $\left( 1+\sqrt{\rho_{\mathrm{anc}}/\rho_{\mathrm{p}}} \right)^2-1\approx2\sqrt{ P_{\mathrm{anc}}w_{\mathrm{p}}^2/\zeta P_{\mathrm{p}}w_{\mathrm{anc}}^2}$  for $P_{\mathrm{anc}}/P_{\mathrm{p}}\ll1$.  Therefore, a counter-intuitively high enhancement ratio of $10\%$ is expected even when $P_{\mathrm{anc}}$ is as small as $60\,\mathrm{\mu W}$.
In Fig.~\ref{fig5}(b), we further plot the enhancement ratio of the trap potential against $P_{\mathrm{anc}}$ with a fixed $P_{\mathrm{p}}=10\,\mathrm{mW}$. As expected, we found that the enhancement rapidly increases with $P_{\mathrm{anc}}$ when $P_{\mathrm{anc}}<200\,\mathrm{\mu W}$ and then linearly increases with $P_{\mathrm{anc}}$ when $P_{\mathrm{anc}}$ is larger.
In particular, when $P_{\mathrm{anc}}=1.3\,\mathrm{mW}$, the trap depth affected by ancillary dipole beam is twice as deep as that when there is no ancillary dipole beam. This means that a counter-propagating $1.3\,\mathrm{mW}$ ancillary laser power has the same influence on the dipole trap depth compared with an addition of $10\,\mathrm{mW}$ power to the primary dipole beam.

In a recent study~\cite{Schymik2022}, the loading of single atoms into an optical tweezer was systematically investigated, and the relation between the loading probability and the power of the dipole trap ($P$) was fitted by the expression
\begin{equation}
\eta(P)=\eta_0\frac{\mathrm{erf}\left[\alpha  \left(P-\text{$P_{\mathrm{half}}$}\right)\right]+1}{2},
\label{equ4}
\end{equation}
where $\eta_0$ is the maximally achievable single atom loading rate which is typically around $50\%$, $\mathrm{erf}[\cdot]$ is the error function, $\alpha$ is a fitting coefficient accounting for the width of the transition region, and $P_{\mathrm{half}}$ is the power at which the loading probability is $\eta_0/2$. According to this equation, the loading probability against the primary dipole beam power is fitted, as shown in Fig.~\ref{fig5}(c), with the dots from the experimental results. The parameters $\alpha=310\,\mathrm{mW}^{-1}$ and $P_{\mathrm{half}}=15.1\,\mathrm{mW}$ were determined for the primary dipole trap created by our metalens. Similarly, in Fig.~\ref{fig5}(d),  we fitted the loading probability against the ancillary dipole beam power with a fixed $P_{\mathrm{p}}=10\,\mathrm{mW}$ and obtained the parameters $\alpha=1100\,\mathrm{mW}^{-1}$ and $P_{\mathrm{half}}=18\,\mu\mathrm{W}$.

If the enhancement of the loading probability is merely due to the enhanced optical trap depth, the fitted $P_{\mathrm{half}}$ for the two scenarios should yield the same trap depth. However, we estimated that the trap depth with $P_{\mathrm{anc}}=18\,\mu\mathrm{W}$ is $0.68\,\mathrm{mK}$, while the trap depth is about $0.83\,\mathrm{mK}$ for primary dipole trap with $P_{\mathrm{p}}=15.2\,\mathrm{mW}$ and $P_{\mathrm{anc}}=0$. This indicates that the ancillary dipole beam not only deepens the trap depth but also introduces a new mechanism for trapping atoms with lower trap potentials. One potential explanation is that a standing-wave-like optical lattice trap provides multiple independent local trap minia, so the probability of trapping single atoms increases with the number of minima. We excluded this possibility because we cannot find a 2-atom signal in these data because of the very weak ancillary dipole beam power. The mechanism might be explained as the ancillary dipole beam alters the shape of the dipole trap, which becomes steeper along the $z$-axis and leads to stronger axial confinement force as $F_z=-\frac{\partial U_{dip}}{\partial z}$. It is also possible that the spatially varying AC-Stark shift arising from the standing-wave dipole trap generates interesting effects, such as Sisyphus cooling~\cite{Hilliard2015}, leading to an increased friction force experienced by atoms along the $z$-axis, which worth further investigation in the future.


\subsection{\label{sec:level5}The high-power regime}

The behavior of the atom loading in the high-power regime was systematically investigated by adjusting the polarization of the ancillary dipole beam while keeping the powers of both dipole beam constant.
In each experimental cycle, the angle of the QWP, which the ancillary dipole beam passed through in Fig.~\ref{fig1}, was changed every 6 seconds in steps of $10^{\circ}$ by a motorized rotation stage. Therefore, we can weaken the potential influence of slowly varying environments and obtain more reliable experimental data.
The results are summarized in Fig.~\ref{fig6}, where the horizontal coordinate $\theta$ denotes the angle between the fast axis of the quarter-wave plate and the horizontal direction. Here, the corresponding angles for realizing horizontal (H), right hand circular (R) and left hand circular (L) polarizations of ancillary dipole beam are also denoted on the axes. The probabilities for loading 0, 1, and 2, or more than 2 atoms when $P_{\mathrm{anc}}=0.5$, $2.5$ and $5\,\mathrm{mW}$ are plotted in Figs.~\ref{fig6}(a)-(c). Figures~\ref{fig6}(d)-(f) represent the typical fluorescence traces of the corresponding data marked by shadows in Figs.~\ref{fig6}(a)-(c), respectively.

The primary dipole beam passed through the metalens is polarized to left hand circular polarization, and thus the interference between the primary and ancillary dipole beam is suppressed for $\theta$ around $145.8^{\circ}$ and is most profound at $\theta=53.0^{\circ}$. From the dependence of the loading probabilities on $\theta$ for $P_{\mathrm{anc}}=0.5\,\mathrm{mW}$ (Fig.~\ref{fig6}(a)), the loading of a single atom and two atoms are obviously modulated when changing $\theta$. 
We find that the system tends to capture single atoms when $126.8^{\circ}\le \theta \le 160.7^{\circ}$, which agrees with our prediction that interference contributes to the enhancement of atom trapping. For example, there are only $2.1\%$ probabilities of capturing 2 atoms when $\theta=145.8^{\circ}$, while the probability of trapping 2 atoms can reach more than $20\%$ when $21.2^{\circ}\le \theta \le 63.7^{\circ}$. 

Similarly, trends are observed for $P_{\mathrm{anc}}=2.5\,\mathrm{mW}$ and $P_{\mathrm{anc}}=5\,\mathrm{mW}$, as illustrated in Figs.~\ref{fig6}(b),(e1)-(e2) and Figs.~\ref{fig6}(c),(f1)-(f2). Since the trap depth is deeper in these cases and the atoms might stay in local minima of the modulated trap along the $z$-direction, the fluorescence signals for more than 2 atoms cases are difficult to be precisely distinguished due to the atom-location dependent AC-Stark shift and fluorescence collection efficiency. Thus, all cases where dipole trap capture more than 2 atoms are summarized and plotted, as shown by the purple dots in the figures. We find that when $126.8^{\circ}\le \theta \le 160.7^{\circ}$, the effect of ancillary dipole beam is weak, and the probability of trapping multiple atoms decreases, leading to a dip. The dip in Fig.~\ref{fig6}(c) is narrower than that in Fig.~\ref{fig6}(b), implying that just a small $P_{\mathrm{anc}}$ polarization component is needed for interfering with the primary dipole beam, the ability of dipole trap to capture multiple atoms can be nearly saturated. However, at $21.2^{\circ}\le \theta \le 63.7^{\circ}$, the ability to trap multiple atoms in Figs.~\ref{fig6}(b) and (c) slightly decreases instead of increasing, as in Fig.~\ref{fig6}(a), predicting that the probability of trapping multiple atoms in Fig.~\ref{fig3}(a) will also decrease slowly after reaching saturation when $P_{\mathrm{anc}}$ is too large. The above results confirm that the reason behind the enhancement of the dipole trap is the interference between the primary and ancillary dipole beam.


\section{\label{sec:level6}Conclusion}
In summary, our study demonstrates the effectiveness of using ancillary dipole beam to enhance the performance of trapping atoms in a tight dipole trap. We have shown that even a very small amount of ancillary dipole beam ($100\,\mu\mathrm{W}$) can significantly modify the trap potential, change the loading rate, and extend the lifetime of single atoms. Notably, this enhancement is not merely contributed by the enhanced trap depth, and might also be due to the enhanced confinement of atoms along the axial direction of the beam. Furthermore, within the high-power regime, we observed the efficient loading of multiple atoms for a period of time, providing a potential approach for preparing few-atom ensembles. Our findings suggest that ancillary dipole beam could be used for efficient controlling of the single-atom loading process in dipole traps, and also underscore the prospective benefits of incorporating ancillary dipole beam in trapping experiments, such as a reduced power requirement for future single-atom array experiments.

\section{\label{sec:level7}acknowledgments}
This work was funded by the National Key R\&D Program (Grant No.~2021YFF0603701), and the National Natural Science Foundation of China (Grants U21A20433, U21A6006, and 11922411). This work was also supported by the Fundamental Research Funds for the Central Universities and USTC Research Funds of the Double First-Class Initiative. K.H. acknowledges the CAS Project for Young Scientists in Basic Research (Grant No.YSBR-049), the National Natural Science Foundation of China (Grant No. 12134013), the National Key Research and Development Program of China (No. 2022YFB3607300), and the CAS Pioneer Hundred Talents Program. The numerical calculations in this paper we performed on the supercomputing system in the Supercomputing Center of the University of Science and Technology of China. This work was partially carried out at the USTC Center for Micro and Nanoscale Research and Fabrication.

\bibliographystyle{Zou}
\nocite{*}
\bibliography{reference}

\begin{thebibliography}{50}%
\makeatletter
\providecommand \@ifxundefined [1]{%
 \@ifx{#1\undefined}
}%
\providecommand \@ifnum [1]{%
 \ifnum #1\expandafter \@firstoftwo
 \else \expandafter \@secondoftwo
 \fi
}%
\providecommand \@ifx [1]{%
 \ifx #1\expandafter \@firstoftwo
 \else \expandafter \@secondoftwo
 \fi
}%
\providecommand \natexlab [1]{#1}%
\providecommand \enquote  [1]{``#1''}%
\providecommand \bibnamefont  [1]{#1}%
\providecommand \bibfnamefont [1]{#1}%
\providecommand \citenamefont [1]{#1}%
\providecommand \href@noop [0]{\@secondoftwo}%
\providecommand \href [0]{\begingroup \@sanitize@url \@href}%
\providecommand \@href[1]{\@@startlink{#1}\@@href}%
\providecommand \@@href[1]{\endgroup#1\@@endlink}%
\providecommand \@sanitize@url [0]{\catcode `\\12\catcode `\$12\catcode
  `\&12\catcode `\#12\catcode `\^12\catcode `\_12\catcode `\%12\relax}%
\providecommand \@@startlink[1]{}%
\providecommand \@@endlink[0]{}%
\providecommand \url  [0]{\begingroup\@sanitize@url \@url }%
\providecommand \@url [1]{\endgroup\@href {#1}{\urlprefix }}%
\providecommand \urlprefix  [0]{URL }%
\providecommand \Eprint [0]{\href }%
\providecommand \doibase [0]{http://dx.doi.org/}%
\providecommand \selectlanguage [0]{\@gobble}%
\providecommand \bibinfo  [0]{\@secondoftwo}%
\providecommand \bibfield  [0]{\@secondoftwo}%
\providecommand \translation [1]{[#1]}%
\providecommand \BibitemOpen [0]{}%
\providecommand \bibitemStop [0]{}%
\providecommand \bibitemNoStop [0]{.\EOS\space}%
\providecommand \EOS [0]{\spacefactor3000\relax}%
\providecommand \BibitemShut  [1]{\csname bibitem#1\endcsname}%
\let\auto@bib@innerbib\@empty
\bibitem [{\citenamefont {Kaufman}\ \emph {et~al.}(2014)\citenamefont
  {Kaufman}, \citenamefont {Lester}, \citenamefont {Reynolds}, \citenamefont
  {Wall}, \citenamefont {Foss-Feig}, \citenamefont {Hazzard}, \citenamefont
  {Rey},\ and\ \citenamefont {Regal}}]{Kaufman2014}%
  \BibitemOpen
  \bibfield  {author} {\bibinfo {author} {\bibfnamefont {A.~M.}\ \bibnamefont
  {Kaufman}}, \bibinfo {author} {\bibfnamefont {B.~J.}\ \bibnamefont {Lester}},
  \bibinfo {author} {\bibfnamefont {C.~M.}\ \bibnamefont {Reynolds}}, \bibinfo
  {author} {\bibfnamefont {M.~L.}\ \bibnamefont {Wall}}, \bibinfo {author}
  {\bibfnamefont {M.}~\bibnamefont {Foss-Feig}}, \bibinfo {author}
  {\bibfnamefont {K.~R.~A.}\ \bibnamefont {Hazzard}}, \bibinfo {author}
  {\bibfnamefont {A.~M.}\ \bibnamefont {Rey}}, \ and\ \bibinfo {author}
  {\bibfnamefont {C.~A.}\ \bibnamefont {Regal}},\ }\bibfield  {title} {\enquote
  {\bibinfo {title} {{Two-particle quantum interference in tunnel-coupled
  optical tweezers}},}\ }\href {\doibase 10.1126/science.1250057} {\bibfield
  {journal} {\bibinfo  {journal} {Science}\ }\textbf {\bibinfo {volume}
  {345}},\ \bibinfo {pages} {306} (\bibinfo {year} {2014})}\BibitemShut
  {NoStop}%
\bibitem [{\citenamefont {Brown}\ \emph {et~al.}(2023)\citenamefont {Brown},
  \citenamefont {Muleady}, \citenamefont {Dworschack}, \citenamefont
  {Lewis-Swan}, \citenamefont {Rey}, \citenamefont {Romero-Isart},\ and\
  \citenamefont {Regal}}]{Brown2023}%
  \BibitemOpen
  \bibfield  {author} {\bibinfo {author} {\bibfnamefont {M.~O.}\ \bibnamefont
  {Brown}}, \bibinfo {author} {\bibfnamefont {S.~R.}\ \bibnamefont {Muleady}},
  \bibinfo {author} {\bibfnamefont {W.~J.}\ \bibnamefont {Dworschack}},
  \bibinfo {author} {\bibfnamefont {R.~J.}\ \bibnamefont {Lewis-Swan}},
  \bibinfo {author} {\bibfnamefont {A.~M.}\ \bibnamefont {Rey}}, \bibinfo
  {author} {\bibfnamefont {O.}~\bibnamefont {Romero-Isart}}, \ and\ \bibinfo
  {author} {\bibfnamefont {C.~A.}\ \bibnamefont {Regal}},\ }\bibfield  {title}
  {\enquote {\bibinfo {title} {{Time-of-flight quantum tomography of an atom in
  an optical tweezer}},}\ }\href {\doibase 10.1038/s41567-022-01890-8}
  {\bibfield  {journal} {\bibinfo  {journal} {Nature Physics}\ }\textbf
  {\bibinfo {volume} {19}},\ \bibinfo {pages} {569} (\bibinfo {year}
  {2023})}\BibitemShut {NoStop}%
\bibitem [{\citenamefont {Tey}\ \emph {et~al.}(2008)\citenamefont {Tey},
  \citenamefont {Chen}, \citenamefont {Aljunid}, \citenamefont {Chng},
  \citenamefont {Huber}, \citenamefont {Maslennikov},\ and\ \citenamefont
  {Kurtsiefer}}]{Tey2008}%
  \BibitemOpen
  \bibfield  {author} {\bibinfo {author} {\bibfnamefont {M.~K.}\ \bibnamefont
  {Tey}}, \bibinfo {author} {\bibfnamefont {Z.}~\bibnamefont {Chen}}, \bibinfo
  {author} {\bibfnamefont {S.~A.}\ \bibnamefont {Aljunid}}, \bibinfo {author}
  {\bibfnamefont {B.}~\bibnamefont {Chng}}, \bibinfo {author} {\bibfnamefont
  {F.}~\bibnamefont {Huber}}, \bibinfo {author} {\bibfnamefont
  {G.}~\bibnamefont {Maslennikov}}, \ and\ \bibinfo {author} {\bibfnamefont
  {C.}~\bibnamefont {Kurtsiefer}},\ }\bibfield  {title} {\enquote {\bibinfo
  {title} {{Strong interaction between light and a single trapped atom without
  the need for a cavity}},}\ }\href {\doibase 10.1038/nphys1096} {\bibfield
  {journal} {\bibinfo  {journal} {Nature Physics}\ }\textbf {\bibinfo {volume}
  {4}},\ \bibinfo {pages} {924} (\bibinfo {year} {2008})}\BibitemShut {NoStop}%
\bibitem [{\citenamefont {O'Shea}\ \emph {et~al.}(2013)\citenamefont {O'Shea},
  \citenamefont {Junge}, \citenamefont {Volz},\ and\ \citenamefont
  {Rauschenbeutel}}]{OShea2013}%
  \BibitemOpen
  \bibfield  {author} {\bibinfo {author} {\bibfnamefont {D.}~\bibnamefont
  {O'Shea}}, \bibinfo {author} {\bibfnamefont {C.}~\bibnamefont {Junge}},
  \bibinfo {author} {\bibfnamefont {J.}~\bibnamefont {Volz}}, \ and\ \bibinfo
  {author} {\bibfnamefont {A.}~\bibnamefont {Rauschenbeutel}},\ }\bibfield
  {title} {\enquote {\bibinfo {title} {{Fiber-Optical Switch Controlled by a
  Single Atom}},}\ }\href {\doibase 10.1103/PhysRevLett.111.193601} {\bibfield
  {journal} {\bibinfo  {journal} {Physical Review Letters}\ }\textbf {\bibinfo
  {volume} {111}},\ \bibinfo {pages} {193601} (\bibinfo {year}
  {2013})}\BibitemShut {NoStop}%
\bibitem [{\citenamefont {Chin}\ \emph {et~al.}(2017)\citenamefont {Chin},
  \citenamefont {Steiner},\ and\ \citenamefont {Kurtsiefer}}]{Chin2017}%
  \BibitemOpen
  \bibfield  {author} {\bibinfo {author} {\bibfnamefont {Y.-S.}\ \bibnamefont
  {Chin}}, \bibinfo {author} {\bibfnamefont {M.}~\bibnamefont {Steiner}}, \
  and\ \bibinfo {author} {\bibfnamefont {C.}~\bibnamefont {Kurtsiefer}},\
  }\bibfield  {title} {\enquote {\bibinfo {title} {{Nonlinear photon-atom
  coupling with 4Pi microscopy}},}\ }\href {\doibase
  10.1038/s41467-017-01495-3} {\bibfield  {journal} {\bibinfo  {journal}
  {Nature Communications}\ }\textbf {\bibinfo {volume} {8}},\ \bibinfo {pages}
  {1200} (\bibinfo {year} {2017})}\BibitemShut {NoStop}%
\bibitem [{\citenamefont {Deist}\ \emph {et~al.}(2022)\citenamefont {Deist},
  \citenamefont {Lu}, \citenamefont {Ho}, \citenamefont {Pasha}, \citenamefont
  {Zeiher}, \citenamefont {Yan},\ and\ \citenamefont
  {Stamper-Kurn}}]{Deist2022}%
  \BibitemOpen
  \bibfield  {author} {\bibinfo {author} {\bibfnamefont {E.}~\bibnamefont
  {Deist}}, \bibinfo {author} {\bibfnamefont {Y.-H.}\ \bibnamefont {Lu}},
  \bibinfo {author} {\bibfnamefont {J.}~\bibnamefont {Ho}}, \bibinfo {author}
  {\bibfnamefont {M.~K.}\ \bibnamefont {Pasha}}, \bibinfo {author}
  {\bibfnamefont {J.}~\bibnamefont {Zeiher}}, \bibinfo {author} {\bibfnamefont
  {Z.}~\bibnamefont {Yan}}, \ and\ \bibinfo {author} {\bibfnamefont {D.~M.}\
  \bibnamefont {Stamper-Kurn}},\ }\bibfield  {title} {\enquote {\bibinfo
  {title} {{Mid-Circuit Cavity Measurement in a Neutral Atom Array}},}\ }\href
  {\doibase 10.1103/PhysRevLett.129.203602} {\bibfield  {journal} {\bibinfo
  {journal} {Physical Review Letters}\ }\textbf {\bibinfo {volume} {129}},\
  \bibinfo {pages} {203602} (\bibinfo {year} {2022})}\BibitemShut {NoStop}%
\bibitem [{\citenamefont {Liu}\ \emph {et~al.}(2023)\citenamefont {Liu},
  \citenamefont {Wang}, \citenamefont {Yang}, \citenamefont {Wang},
  \citenamefont {Fan}, \citenamefont {Guan}, \citenamefont {Li}, \citenamefont
  {Zhang},\ and\ \citenamefont {Zhang}}]{Liu2023}%
  \BibitemOpen
  \bibfield  {author} {\bibinfo {author} {\bibfnamefont {Y.}~\bibnamefont
  {Liu}}, \bibinfo {author} {\bibfnamefont {Z.}~\bibnamefont {Wang}}, \bibinfo
  {author} {\bibfnamefont {P.}~\bibnamefont {Yang}}, \bibinfo {author}
  {\bibfnamefont {Q.}~\bibnamefont {Wang}}, \bibinfo {author} {\bibfnamefont
  {Q.}~\bibnamefont {Fan}}, \bibinfo {author} {\bibfnamefont {S.}~\bibnamefont
  {Guan}}, \bibinfo {author} {\bibfnamefont {G.}~\bibnamefont {Li}}, \bibinfo
  {author} {\bibfnamefont {P.}~\bibnamefont {Zhang}}, \ and\ \bibinfo {author}
  {\bibfnamefont {T.}~\bibnamefont {Zhang}},\ }\bibfield  {title} {\enquote
  {\bibinfo {title} {{Realization of Strong Coupling between Deterministic
  Single-Atom Arrays and a High-Finesse Miniature Optical Cavity}},}\ }\href
  {\doibase 10.1103/PhysRevLett.130.173601} {\bibfield  {journal} {\bibinfo
  {journal} {Physical Review Letters}\ }\textbf {\bibinfo {volume} {130}},\
  \bibinfo {pages} {173601} (\bibinfo {year} {2023})}\BibitemShut {NoStop}%
\bibitem [{\citenamefont {Will}\ \emph {et~al.}(2021)\citenamefont {Will},
  \citenamefont {Masters}, \citenamefont {Rauschenbeutel}, \citenamefont
  {Scheucher},\ and\ \citenamefont {Volz}}]{Will2021}%
  \BibitemOpen
  \bibfield  {author} {\bibinfo {author} {\bibfnamefont {E.}~\bibnamefont
  {Will}}, \bibinfo {author} {\bibfnamefont {L.}~\bibnamefont {Masters}},
  \bibinfo {author} {\bibfnamefont {A.}~\bibnamefont {Rauschenbeutel}},
  \bibinfo {author} {\bibfnamefont {M.}~\bibnamefont {Scheucher}}, \ and\
  \bibinfo {author} {\bibfnamefont {J.}~\bibnamefont {Volz}},\ }\bibfield
  {title} {\enquote {\bibinfo {title} {{Coupling a Single Trapped Atom to a
  Whispering-Gallery-Mode Microresonator}},}\ }\href {\doibase
  10.1103/PhysRevLett.126.233602} {\bibfield  {journal} {\bibinfo  {journal}
  {Physical Review Letters}\ }\textbf {\bibinfo {volume} {126}},\ \bibinfo
  {pages} {233602} (\bibinfo {year} {2021})}\BibitemShut {NoStop}%
\bibitem [{\citenamefont {Zhou}\ \emph {et~al.}(2023)\citenamefont {Zhou},
  \citenamefont {Tamura}, \citenamefont {Chang},\ and\ \citenamefont
  {Hung}}]{Zhou2023}%
  \BibitemOpen
  \bibfield  {author} {\bibinfo {author} {\bibfnamefont {X.}~\bibnamefont
  {Zhou}}, \bibinfo {author} {\bibfnamefont {H.}~\bibnamefont {Tamura}},
  \bibinfo {author} {\bibfnamefont {T.-H.}\ \bibnamefont {Chang}}, \ and\
  \bibinfo {author} {\bibfnamefont {C.-L.}\ \bibnamefont {Hung}},\ }\bibfield
  {title} {\enquote {\bibinfo {title} {{Coupling Single Atoms to a Nanophotonic
  Whispering-Gallery-Mode Resonator via Optical Guiding}},}\ }\href {\doibase
  10.1103/PhysRevLett.130.103601} {\bibfield  {journal} {\bibinfo  {journal}
  {Physical Review Letters}\ }\textbf {\bibinfo {volume} {130}},\ \bibinfo
  {pages} {103601} (\bibinfo {year} {2023})}\BibitemShut {NoStop}%
\bibitem [{\citenamefont {Saffman}(2019)}]{Saffman2019}%
  \BibitemOpen
  \bibfield  {author} {\bibinfo {author} {\bibfnamefont {M.}~\bibnamefont
  {Saffman}},\ }\bibfield  {title} {\enquote {\bibinfo {title} {{Quantum
  computing with neutral atoms}},}\ }\href {\doibase 10.1093/nsr/nwy088}
  {\bibfield  {journal} {\bibinfo  {journal} {National Science Review}\
  }\textbf {\bibinfo {volume} {6}},\ \bibinfo {pages} {24} (\bibinfo {year}
  {2019})}\BibitemShut {NoStop}%
\bibitem [{\citenamefont {Browaeys}\ and\ \citenamefont
  {Lahaye}(2020)}]{Browaeys2020}%
  \BibitemOpen
  \bibfield  {author} {\bibinfo {author} {\bibfnamefont {A.}~\bibnamefont
  {Browaeys}}\ and\ \bibinfo {author} {\bibfnamefont {T.}~\bibnamefont
  {Lahaye}},\ }\bibfield  {title} {\enquote {\bibinfo {title} {{Many-body
  physics with individually controlled Rydberg atoms}},}\ }\href {\doibase
  10.1038/s41567-019-0733-z} {\bibfield  {journal} {\bibinfo  {journal} {Nature
  Physics}\ }\textbf {\bibinfo {volume} {16}},\ \bibinfo {pages} {132}
  (\bibinfo {year} {2020})}\BibitemShut {NoStop}%
\bibitem [{\citenamefont {Morgado}\ and\ \citenamefont
  {Whitlock}(2021)}]{Morgado2021}%
  \BibitemOpen
  \bibfield  {author} {\bibinfo {author} {\bibfnamefont {M.}~\bibnamefont
  {Morgado}}\ and\ \bibinfo {author} {\bibfnamefont {S.}~\bibnamefont
  {Whitlock}},\ }\bibfield  {title} {\enquote {\bibinfo {title} {{Quantum
  simulation and computing with Rydberg-interacting qubits}},}\ }\href
  {\doibase 10.1116/5.0036562} {\bibfield  {journal} {\bibinfo  {journal} {AVS
  Quantum Science}\ }\textbf {\bibinfo {volume} {3}},\ \bibinfo {pages}
  {023501} (\bibinfo {year} {2021})}\BibitemShut {NoStop}%
\bibitem [{\citenamefont {Wu}\ \emph {et~al.}(2021)\citenamefont {Wu},
  \citenamefont {Liang}, \citenamefont {Tian}, \citenamefont {Yang},
  \citenamefont {Chen}, \citenamefont {Liu}, \citenamefont {Tey},\ and\
  \citenamefont {You}}]{Wu2021}%
  \BibitemOpen
  \bibfield  {author} {\bibinfo {author} {\bibfnamefont {X.}~\bibnamefont
  {Wu}}, \bibinfo {author} {\bibfnamefont {X.}~\bibnamefont {Liang}}, \bibinfo
  {author} {\bibfnamefont {Y.}~\bibnamefont {Tian}}, \bibinfo {author}
  {\bibfnamefont {F.}~\bibnamefont {Yang}}, \bibinfo {author} {\bibfnamefont
  {C.}~\bibnamefont {Chen}}, \bibinfo {author} {\bibfnamefont {Y.-C.}\
  \bibnamefont {Liu}}, \bibinfo {author} {\bibfnamefont {M.~K.}\ \bibnamefont
  {Tey}}, \ and\ \bibinfo {author} {\bibfnamefont {L.}~\bibnamefont {You}},\
  }\bibfield  {title} {\enquote {\bibinfo {title} {{A concise review of Rydberg
  atom based quantum computation and quantum simulation*}},}\ }\href {\doibase
  10.1088/1674-1056/abd76f} {\bibfield  {journal} {\bibinfo  {journal} {Chinese
  Physics B}\ }\textbf {\bibinfo {volume} {30}},\ \bibinfo {pages} {020305}
  (\bibinfo {year} {2021})}\BibitemShut {NoStop}%
\bibitem [{\citenamefont {Grimm}\ \emph {et~al.}(2000)\citenamefont {Grimm},
  \citenamefont {Weidem{\"{u}}ller},\ and\ \citenamefont
  {Ovchinnikov}}]{Grimm2000}%
  \BibitemOpen
  \bibfield  {author} {\bibinfo {author} {\bibfnamefont {R.}~\bibnamefont
  {Grimm}}, \bibinfo {author} {\bibfnamefont {M.}~\bibnamefont
  {Weidem{\"{u}}ller}}, \ and\ \bibinfo {author} {\bibfnamefont {Y.~B.}\
  \bibnamefont {Ovchinnikov}},\ }\bibfield  {title} {\enquote {\bibinfo {title}
  {{Optical Dipole Traps for Neutral Atoms}},}\ }in\ \href {\doibase
  10.1016/S1049-250X(08)60186-X} {\emph {\bibinfo {booktitle} {Advances in
  Atomic, Molecular, and Optical Physics}}},\ \bibinfo {series and number}
  {\bibinfo {number} {December}}\ (\bibinfo {year} {2000})\ pp.\ \bibinfo
  {pages} {95--170}\BibitemShut {NoStop}%
\bibitem [{\citenamefont {Kuppens}\ \emph {et~al.}(2000)\citenamefont
  {Kuppens}, \citenamefont {Corwin}, \citenamefont {Miller}, \citenamefont
  {Chupp},\ and\ \citenamefont {Wieman}}]{Kuppens2000}%
  \BibitemOpen
  \bibfield  {author} {\bibinfo {author} {\bibfnamefont {S.~J.~M.}\
  \bibnamefont {Kuppens}}, \bibinfo {author} {\bibfnamefont {K.~L.}\
  \bibnamefont {Corwin}}, \bibinfo {author} {\bibfnamefont {K.~W.}\
  \bibnamefont {Miller}}, \bibinfo {author} {\bibfnamefont {T.~E.}\
  \bibnamefont {Chupp}}, \ and\ \bibinfo {author} {\bibfnamefont {C.~E.}\
  \bibnamefont {Wieman}},\ }\bibfield  {title} {\enquote {\bibinfo {title}
  {{Loading an optical dipole trap}},}\ }\href {\doibase
  10.1103/PhysRevA.62.013406} {\bibfield  {journal} {\bibinfo  {journal}
  {Physical Review A}\ }\textbf {\bibinfo {volume} {62}},\ \bibinfo {pages}
  {013406} (\bibinfo {year} {2000})}\BibitemShut {NoStop}%
\bibitem [{\citenamefont {Li}\ \emph {et~al.}(2020)\citenamefont {Li},
  \citenamefont {Li}, \citenamefont {Yang}, \citenamefont {Wang}, \citenamefont
  {Zhang},\ and\ \citenamefont {Zhang}}]{Haokang2020}%
  \BibitemOpen
  \bibfield  {author} {\bibinfo {author} {\bibfnamefont {S.}~\bibnamefont
  {Li}}, \bibinfo {author} {\bibfnamefont {G.}~\bibnamefont {Li}}, \bibinfo
  {author} {\bibfnamefont {P.}~\bibnamefont {Yang}}, \bibinfo {author}
  {\bibfnamefont {Z.}~\bibnamefont {Wang}}, \bibinfo {author} {\bibfnamefont
  {P.}~\bibnamefont {Zhang}}, \ and\ \bibinfo {author} {\bibfnamefont
  {T.}~\bibnamefont {Zhang}},\ }\bibfield  {title} {\enquote {\bibinfo {title}
  {{Versatile objectives with NA = 0.55 and NA = 0.78 for cold-atom
  experiments}},}\ }\href {\doibase 10.1364/OE.408945} {\bibfield  {journal}
  {\bibinfo  {journal} {Optics Express}\ }\textbf {\bibinfo {volume} {28}},\
  \bibinfo {pages} {36122} (\bibinfo {year} {2020})}\BibitemShut {NoStop}%
\bibitem [{\citenamefont {Schlosser}\ \emph {et~al.}(2001)\citenamefont
  {Schlosser}, \citenamefont {Reymond}, \citenamefont {Protsenko},\ and\
  \citenamefont {Grangier}}]{Schlosser2001}%
  \BibitemOpen
  \bibfield  {author} {\bibinfo {author} {\bibfnamefont {N.}~\bibnamefont
  {Schlosser}}, \bibinfo {author} {\bibfnamefont {G.}~\bibnamefont {Reymond}},
  \bibinfo {author} {\bibfnamefont {I.}~\bibnamefont {Protsenko}}, \ and\
  \bibinfo {author} {\bibfnamefont {P.}~\bibnamefont {Grangier}},\ }\bibfield
  {title} {\enquote {\bibinfo {title} {{Sub-poissonian loading of single atoms
  in a microscopic dipole trap}},}\ }\href {\doibase 10.1038/35082512}
  {\bibfield  {journal} {\bibinfo  {journal} {Nature}\ }\textbf {\bibinfo
  {volume} {411}},\ \bibinfo {pages} {1024} (\bibinfo {year}
  {2001})}\BibitemShut {NoStop}%
\bibitem [{\citenamefont {Schlosser}\ \emph {et~al.}(2002)\citenamefont
  {Schlosser}, \citenamefont {Reymond},\ and\ \citenamefont
  {Grangier}}]{Schlosser2002}%
  \BibitemOpen
  \bibfield  {author} {\bibinfo {author} {\bibfnamefont {N.}~\bibnamefont
  {Schlosser}}, \bibinfo {author} {\bibfnamefont {G.}~\bibnamefont {Reymond}},
  \ and\ \bibinfo {author} {\bibfnamefont {P.}~\bibnamefont {Grangier}},\
  }\bibfield  {title} {\enquote {\bibinfo {title} {{Collisional Blockade in
  Microscopic Optical Dipole Traps}},}\ }\href {\doibase
  10.1103/PhysRevLett.89.023005} {\bibfield  {journal} {\bibinfo  {journal}
  {Physical Review Letters}\ }\textbf {\bibinfo {volume} {89}},\ \bibinfo
  {pages} {023005} (\bibinfo {year} {2002})}\BibitemShut {NoStop}%
\bibitem [{\citenamefont {Fung}\ and\ \citenamefont
  {Andersen}(2015)}]{Fung2015}%
  \BibitemOpen
  \bibfield  {author} {\bibinfo {author} {\bibfnamefont {Y.~H.}\ \bibnamefont
  {Fung}}\ and\ \bibinfo {author} {\bibfnamefont {M.~F.}\ \bibnamefont
  {Andersen}},\ }\bibfield  {title} {\enquote {\bibinfo {title} {{Efficient
  collisional blockade loading of a single atom into a tight microtrap}},}\
  }\href {\doibase 10.1088/1367-2630/17/7/073011} {\bibfield  {journal}
  {\bibinfo  {journal} {New Journal of Physics}\ }\textbf {\bibinfo {volume}
  {17}},\ \bibinfo {pages} {073011} (\bibinfo {year} {2015})}\BibitemShut
  {NoStop}%
\bibitem [{\citenamefont {Barredo}\ \emph {et~al.}(2016)\citenamefont
  {Barredo}, \citenamefont {de~L{\'{e}}s{\'{e}}leuc}, \citenamefont {Lienhard},
  \citenamefont {Lahaye},\ and\ \citenamefont {Browaeys}}]{Barredo2016}%
  \BibitemOpen
  \bibfield  {author} {\bibinfo {author} {\bibfnamefont {D.}~\bibnamefont
  {Barredo}}, \bibinfo {author} {\bibfnamefont {S.}~\bibnamefont
  {de~L{\'{e}}s{\'{e}}leuc}}, \bibinfo {author} {\bibfnamefont
  {V.}~\bibnamefont {Lienhard}}, \bibinfo {author} {\bibfnamefont
  {T.}~\bibnamefont {Lahaye}}, \ and\ \bibinfo {author} {\bibfnamefont
  {A.}~\bibnamefont {Browaeys}},\ }\bibfield  {title} {\enquote {\bibinfo
  {title} {{An atom-by-atom assembler of defect-free arbitrary two-dimensional
  atomic arrays}},}\ }\href {\doibase 10.1126/science.aah3778} {\bibfield
  {journal} {\bibinfo  {journal} {Science}\ }\textbf {\bibinfo {volume}
  {354}},\ \bibinfo {pages} {1021} (\bibinfo {year} {2016})}\BibitemShut
  {NoStop}%
\bibitem [{\citenamefont {Endres}\ \emph {et~al.}(2016)\citenamefont {Endres},
  \citenamefont {Bernien}, \citenamefont {Keesling}, \citenamefont {Levine},
  \citenamefont {Anschuetz}, \citenamefont {Krajenbrink}, \citenamefont
  {Senko}, \citenamefont {Vuletic}, \citenamefont {Greiner},\ and\
  \citenamefont {Lukin}}]{Endres2016}%
  \BibitemOpen
  \bibfield  {author} {\bibinfo {author} {\bibfnamefont {M.}~\bibnamefont
  {Endres}}, \bibinfo {author} {\bibfnamefont {H.}~\bibnamefont {Bernien}},
  \bibinfo {author} {\bibfnamefont {A.}~\bibnamefont {Keesling}}, \bibinfo
  {author} {\bibfnamefont {H.}~\bibnamefont {Levine}}, \bibinfo {author}
  {\bibfnamefont {E.~R.}\ \bibnamefont {Anschuetz}}, \bibinfo {author}
  {\bibfnamefont {A.}~\bibnamefont {Krajenbrink}}, \bibinfo {author}
  {\bibfnamefont {C.}~\bibnamefont {Senko}}, \bibinfo {author} {\bibfnamefont
  {V.}~\bibnamefont {Vuletic}}, \bibinfo {author} {\bibfnamefont
  {M.}~\bibnamefont {Greiner}}, \ and\ \bibinfo {author} {\bibfnamefont
  {M.~D.}\ \bibnamefont {Lukin}},\ }\bibfield  {title} {\enquote {\bibinfo
  {title} {{Atom-by-atom assembly of defect-free one-dimensional cold atom
  arrays}},}\ }\href {\doibase 10.1126/science.aah3752} {\bibfield  {journal}
  {\bibinfo  {journal} {Science}\ }\textbf {\bibinfo {volume} {354}},\ \bibinfo
  {pages} {1024} (\bibinfo {year} {2016})}\BibitemShut {NoStop}%
\bibitem [{\citenamefont {Bernien}\ \emph {et~al.}(2017)\citenamefont
  {Bernien}, \citenamefont {Schwartz}, \citenamefont {Keesling}, \citenamefont
  {Levine}, \citenamefont {Omran}, \citenamefont {Pichler}, \citenamefont
  {Choi}, \citenamefont {Zibrov}, \citenamefont {Endres}, \citenamefont
  {Greiner}, \citenamefont {Vuletic},\ and\ \citenamefont
  {Lukin}}]{Bernien2017}%
  \BibitemOpen
  \bibfield  {author} {\bibinfo {author} {\bibfnamefont {H.}~\bibnamefont
  {Bernien}}, \bibinfo {author} {\bibfnamefont {S.}~\bibnamefont {Schwartz}},
  \bibinfo {author} {\bibfnamefont {A.}~\bibnamefont {Keesling}}, \bibinfo
  {author} {\bibfnamefont {H.}~\bibnamefont {Levine}}, \bibinfo {author}
  {\bibfnamefont {A.}~\bibnamefont {Omran}}, \bibinfo {author} {\bibfnamefont
  {H.}~\bibnamefont {Pichler}}, \bibinfo {author} {\bibfnamefont
  {S.}~\bibnamefont {Choi}}, \bibinfo {author} {\bibfnamefont {A.~S.}\
  \bibnamefont {Zibrov}}, \bibinfo {author} {\bibfnamefont {M.}~\bibnamefont
  {Endres}}, \bibinfo {author} {\bibfnamefont {M.}~\bibnamefont {Greiner}},
  \bibinfo {author} {\bibfnamefont {V.}~\bibnamefont {Vuletic}}, \ and\
  \bibinfo {author} {\bibfnamefont {M.~D.}\ \bibnamefont {Lukin}},\ }\bibfield
  {title} {\enquote {\bibinfo {title} {{Probing many-body dynamics on a 51-atom
  quantum simulator}},}\ }\href {\doibase 10.1038/nature24622} {\bibfield
  {journal} {\bibinfo  {journal} {Nature}\ }\textbf {\bibinfo {volume} {551}},\
  \bibinfo {pages} {579} (\bibinfo {year} {2017})}\BibitemShut {NoStop}%
\bibitem [{\citenamefont {Kaufman}\ and\ \citenamefont
  {Ni}(2021)}]{Kaufman2021}%
  \BibitemOpen
  \bibfield  {author} {\bibinfo {author} {\bibfnamefont {A.~M.}\ \bibnamefont
  {Kaufman}}\ and\ \bibinfo {author} {\bibfnamefont {K.~K.}\ \bibnamefont
  {Ni}},\ }\bibfield  {title} {\enquote {\bibinfo {title} {{Quantum science
  with optical tweezer arrays of ultracold atoms and molecules}},}\ }\href
  {\doibase 10.1038/s41567-021-01357-2} {\bibfield  {journal} {\bibinfo
  {journal} {Nature Physics}\ }\textbf {\bibinfo {volume} {17}},\ \bibinfo
  {pages} {1324} (\bibinfo {year} {2021})}\BibitemShut {NoStop}%
\bibitem [{\citenamefont {Ebadi}\ \emph {et~al.}(2021)\citenamefont {Ebadi},
  \citenamefont {Wang}, \citenamefont {Levine}, \citenamefont {Keesling},
  \citenamefont {Semeghini}, \citenamefont {Omran}, \citenamefont {Bluvstein},
  \citenamefont {Samajdar}, \citenamefont {Pichler}, \citenamefont {Ho},
  \citenamefont {Choi}, \citenamefont {Sachdev}, \citenamefont {Greiner},
  \citenamefont {Vuleti{\'{c}}},\ and\ \citenamefont {Lukin}}]{Ebadi2021}%
  \BibitemOpen
  \bibfield  {author} {\bibinfo {author} {\bibfnamefont {S.}~\bibnamefont
  {Ebadi}}, \bibinfo {author} {\bibfnamefont {T.~T.}\ \bibnamefont {Wang}},
  \bibinfo {author} {\bibfnamefont {H.}~\bibnamefont {Levine}}, \bibinfo
  {author} {\bibfnamefont {A.}~\bibnamefont {Keesling}}, \bibinfo {author}
  {\bibfnamefont {G.}~\bibnamefont {Semeghini}}, \bibinfo {author}
  {\bibfnamefont {A.}~\bibnamefont {Omran}}, \bibinfo {author} {\bibfnamefont
  {D.}~\bibnamefont {Bluvstein}}, \bibinfo {author} {\bibfnamefont
  {R.}~\bibnamefont {Samajdar}}, \bibinfo {author} {\bibfnamefont
  {H.}~\bibnamefont {Pichler}}, \bibinfo {author} {\bibfnamefont {W.~W.}\
  \bibnamefont {Ho}}, \bibinfo {author} {\bibfnamefont {S.}~\bibnamefont
  {Choi}}, \bibinfo {author} {\bibfnamefont {S.}~\bibnamefont {Sachdev}},
  \bibinfo {author} {\bibfnamefont {M.}~\bibnamefont {Greiner}}, \bibinfo
  {author} {\bibfnamefont {V.}~\bibnamefont {Vuleti{\'{c}}}}, \ and\ \bibinfo
  {author} {\bibfnamefont {M.~D.}\ \bibnamefont {Lukin}},\ }\bibfield  {title}
  {\enquote {\bibinfo {title} {{Quantum phases of matter on a 256-atom
  programmable quantum simulator}},}\ }\href {\doibase
  10.1038/s41586-021-03582-4} {\bibfield  {journal} {\bibinfo  {journal}
  {Nature}\ }\textbf {\bibinfo {volume} {595}},\ \bibinfo {pages} {227}
  (\bibinfo {year} {2021})}\BibitemShut {NoStop}%
\bibitem [{\citenamefont {Sheng}\ \emph {et~al.}(2022)\citenamefont {Sheng},
  \citenamefont {Hou}, \citenamefont {He}, \citenamefont {Wang}, \citenamefont
  {Guo}, \citenamefont {Zhuang}, \citenamefont {Mamat}, \citenamefont {Xu},
  \citenamefont {Liu}, \citenamefont {Wang},\ and\ \citenamefont
  {Zhan}}]{Sheng2022}%
  \BibitemOpen
  \bibfield  {author} {\bibinfo {author} {\bibfnamefont {C.}~\bibnamefont
  {Sheng}}, \bibinfo {author} {\bibfnamefont {J.}~\bibnamefont {Hou}}, \bibinfo
  {author} {\bibfnamefont {X.}~\bibnamefont {He}}, \bibinfo {author}
  {\bibfnamefont {K.}~\bibnamefont {Wang}}, \bibinfo {author} {\bibfnamefont
  {R.}~\bibnamefont {Guo}}, \bibinfo {author} {\bibfnamefont {J.}~\bibnamefont
  {Zhuang}}, \bibinfo {author} {\bibfnamefont {B.}~\bibnamefont {Mamat}},
  \bibinfo {author} {\bibfnamefont {P.}~\bibnamefont {Xu}}, \bibinfo {author}
  {\bibfnamefont {M.}~\bibnamefont {Liu}}, \bibinfo {author} {\bibfnamefont
  {J.}~\bibnamefont {Wang}}, \ and\ \bibinfo {author} {\bibfnamefont
  {M.}~\bibnamefont {Zhan}},\ }\bibfield  {title} {\enquote {\bibinfo {title}
  {{Defect-Free Arbitrary-Geometry Assembly of Mixed-Species Atom Arrays}},}\
  }\href {\doibase 10.1103/PhysRevLett.128.083202} {\bibfield  {journal}
  {\bibinfo  {journal} {Physical Review Letters}\ }\textbf {\bibinfo {volume}
  {128}},\ \bibinfo {pages} {083202} (\bibinfo {year} {2022})}\BibitemShut
  {NoStop}%
\bibitem [{\citenamefont {Barredo}\ \emph {et~al.}(2018)\citenamefont
  {Barredo}, \citenamefont {Lienhard}, \citenamefont {de~L{\'{e}}s{\'{e}}leuc},
  \citenamefont {Lahaye},\ and\ \citenamefont {Browaeys}}]{Barredo2018}%
  \BibitemOpen
  \bibfield  {author} {\bibinfo {author} {\bibfnamefont {D.}~\bibnamefont
  {Barredo}}, \bibinfo {author} {\bibfnamefont {V.}~\bibnamefont {Lienhard}},
  \bibinfo {author} {\bibfnamefont {S.}~\bibnamefont
  {de~L{\'{e}}s{\'{e}}leuc}}, \bibinfo {author} {\bibfnamefont
  {T.}~\bibnamefont {Lahaye}}, \ and\ \bibinfo {author} {\bibfnamefont
  {A.}~\bibnamefont {Browaeys}},\ }\bibfield  {title} {\enquote {\bibinfo
  {title} {{Synthetic three-dimensional atomic structures assembled atom by
  atom}},}\ }\href {\doibase 10.1038/s41586-018-0450-2} {\bibfield  {journal}
  {\bibinfo  {journal} {Nature}\ }\textbf {\bibinfo {volume} {561}},\ \bibinfo
  {pages} {79} (\bibinfo {year} {2018})}\BibitemShut {NoStop}%
\bibitem [{\citenamefont {Labuhn}\ \emph {et~al.}(2016)\citenamefont {Labuhn},
  \citenamefont {Barredo}, \citenamefont {Ravets}, \citenamefont {{De
  L{\'{e}}s{\'{e}}leuc}}, \citenamefont {Macr{\`{i}}}, \citenamefont {Lahaye},\
  and\ \citenamefont {Browaeys}}]{Labuhn2016}%
  \BibitemOpen
  \bibfield  {author} {\bibinfo {author} {\bibfnamefont {H.}~\bibnamefont
  {Labuhn}}, \bibinfo {author} {\bibfnamefont {D.}~\bibnamefont {Barredo}},
  \bibinfo {author} {\bibfnamefont {S.}~\bibnamefont {Ravets}}, \bibinfo
  {author} {\bibfnamefont {S.}~\bibnamefont {{De L{\'{e}}s{\'{e}}leuc}}},
  \bibinfo {author} {\bibfnamefont {T.}~\bibnamefont {Macr{\`{i}}}}, \bibinfo
  {author} {\bibfnamefont {T.}~\bibnamefont {Lahaye}}, \ and\ \bibinfo {author}
  {\bibfnamefont {A.}~\bibnamefont {Browaeys}},\ }\bibfield  {title} {\enquote
  {\bibinfo {title} {{Tunable two-dimensional arrays of single Rydberg atoms
  for realizing quantum Ising models}},}\ }\href {\doibase 10.1038/nature18274}
  {\bibfield  {journal} {\bibinfo  {journal} {Nature}\ }\textbf {\bibinfo
  {volume} {534}},\ \bibinfo {pages} {667} (\bibinfo {year}
  {2016})}\BibitemShut {NoStop}%
\bibitem [{\citenamefont {Omran}\ \emph {et~al.}(2019)\citenamefont {Omran},
  \citenamefont {Levine}, \citenamefont {Keesling}, \citenamefont {Semeghini},
  \citenamefont {Wang}, \citenamefont {Ebadi}, \citenamefont {Bernien},
  \citenamefont {Zibrov}, \citenamefont {Pichler}, \citenamefont {Choi},
  \citenamefont {Cui}, \citenamefont {Rossignolo}, \citenamefont {Rembold},
  \citenamefont {Montangero}, \citenamefont {Calarco}, \citenamefont {Endres},
  \citenamefont {Greiner}, \citenamefont {Vuleti{\'{c}}},\ and\ \citenamefont
  {Lukin}}]{Omran2019}%
  \BibitemOpen
  \bibfield  {author} {\bibinfo {author} {\bibfnamefont {A.}~\bibnamefont
  {Omran}}, \bibinfo {author} {\bibfnamefont {H.}~\bibnamefont {Levine}},
  \bibinfo {author} {\bibfnamefont {A.}~\bibnamefont {Keesling}}, \bibinfo
  {author} {\bibfnamefont {G.}~\bibnamefont {Semeghini}}, \bibinfo {author}
  {\bibfnamefont {T.~T.}\ \bibnamefont {Wang}}, \bibinfo {author}
  {\bibfnamefont {S.}~\bibnamefont {Ebadi}}, \bibinfo {author} {\bibfnamefont
  {H.}~\bibnamefont {Bernien}}, \bibinfo {author} {\bibfnamefont {A.~S.}\
  \bibnamefont {Zibrov}}, \bibinfo {author} {\bibfnamefont {H.}~\bibnamefont
  {Pichler}}, \bibinfo {author} {\bibfnamefont {S.}~\bibnamefont {Choi}},
  \bibinfo {author} {\bibfnamefont {J.}~\bibnamefont {Cui}}, \bibinfo {author}
  {\bibfnamefont {M.}~\bibnamefont {Rossignolo}}, \bibinfo {author}
  {\bibfnamefont {P.}~\bibnamefont {Rembold}}, \bibinfo {author} {\bibfnamefont
  {S.}~\bibnamefont {Montangero}}, \bibinfo {author} {\bibfnamefont
  {T.}~\bibnamefont {Calarco}}, \bibinfo {author} {\bibfnamefont
  {M.}~\bibnamefont {Endres}}, \bibinfo {author} {\bibfnamefont
  {M.}~\bibnamefont {Greiner}}, \bibinfo {author} {\bibfnamefont
  {V.}~\bibnamefont {Vuleti{\'{c}}}}, \ and\ \bibinfo {author} {\bibfnamefont
  {M.~D.}\ \bibnamefont {Lukin}},\ }\bibfield  {title} {\enquote {\bibinfo
  {title} {{Generation and manipulation of Schr{\"{o}}dinger cat states in
  Rydberg atom arrays}},}\ }\href {\doibase 10.1126/science.aax9743} {\bibfield
   {journal} {\bibinfo  {journal} {Science}\ }\textbf {\bibinfo {volume}
  {365}},\ \bibinfo {pages} {570} (\bibinfo {year} {2019})}\BibitemShut
  {NoStop}%
\bibitem [{\citenamefont {Shi}\ \emph {et~al.}(2022)\citenamefont {Shi},
  \citenamefont {Xu}, \citenamefont {Zhang}, \citenamefont {Ye}, \citenamefont
  {Xiang}, \citenamefont {Liu}, \citenamefont {Wang}, \citenamefont {Su},\ and\
  \citenamefont {Li}}]{Shi2022}%
  \BibitemOpen
  \bibfield  {author} {\bibinfo {author} {\bibfnamefont {S.}~\bibnamefont
  {Shi}}, \bibinfo {author} {\bibfnamefont {B.}~\bibnamefont {Xu}}, \bibinfo
  {author} {\bibfnamefont {K.}~\bibnamefont {Zhang}}, \bibinfo {author}
  {\bibfnamefont {G.~S.}\ \bibnamefont {Ye}}, \bibinfo {author} {\bibfnamefont
  {D.~S.}\ \bibnamefont {Xiang}}, \bibinfo {author} {\bibfnamefont
  {Y.}~\bibnamefont {Liu}}, \bibinfo {author} {\bibfnamefont {J.}~\bibnamefont
  {Wang}}, \bibinfo {author} {\bibfnamefont {D.}~\bibnamefont {Su}}, \ and\
  \bibinfo {author} {\bibfnamefont {L.}~\bibnamefont {Li}},\ }\bibfield
  {title} {\enquote {\bibinfo {title} {{High-fidelity photonic quantum logic
  gate based on near-optimal Rydberg single-photon source}},}\ }\href {\doibase
  10.1038/s41467-022-32083-9} {\bibfield  {journal} {\bibinfo  {journal}
  {Nature Communications}\ }\textbf {\bibinfo {volume} {13}},\ \bibinfo {pages}
  {4454} (\bibinfo {year} {2022})}\BibitemShut {NoStop}%
\bibitem [{\citenamefont {Bluvstein}\ \emph {et~al.}(2022)\citenamefont
  {Bluvstein}, \citenamefont {Levine}, \citenamefont {Semeghini}, \citenamefont
  {Wang}, \citenamefont {Ebadi}, \citenamefont {Kalinowski}, \citenamefont
  {Keesling}, \citenamefont {Maskara}, \citenamefont {Pichler}, \citenamefont
  {Greiner}, \citenamefont {Vuleti{\'{c}}},\ and\ \citenamefont
  {Lukin}}]{Bluvstein2022}%
  \BibitemOpen
  \bibfield  {author} {\bibinfo {author} {\bibfnamefont {D.}~\bibnamefont
  {Bluvstein}}, \bibinfo {author} {\bibfnamefont {H.}~\bibnamefont {Levine}},
  \bibinfo {author} {\bibfnamefont {G.}~\bibnamefont {Semeghini}}, \bibinfo
  {author} {\bibfnamefont {T.~T.}\ \bibnamefont {Wang}}, \bibinfo {author}
  {\bibfnamefont {S.}~\bibnamefont {Ebadi}}, \bibinfo {author} {\bibfnamefont
  {M.}~\bibnamefont {Kalinowski}}, \bibinfo {author} {\bibfnamefont
  {A.}~\bibnamefont {Keesling}}, \bibinfo {author} {\bibfnamefont
  {N.}~\bibnamefont {Maskara}}, \bibinfo {author} {\bibfnamefont
  {H.}~\bibnamefont {Pichler}}, \bibinfo {author} {\bibfnamefont
  {M.}~\bibnamefont {Greiner}}, \bibinfo {author} {\bibfnamefont
  {V.}~\bibnamefont {Vuleti{\'{c}}}}, \ and\ \bibinfo {author} {\bibfnamefont
  {M.~D.}\ \bibnamefont {Lukin}},\ }\bibfield  {title} {\enquote {\bibinfo
  {title} {{A quantum processor based on coherent transport of entangled atom
  arrays}},}\ }\href {\doibase 10.1038/s41586-022-04592-6} {\bibfield
  {journal} {\bibinfo  {journal} {Nature}\ }\textbf {\bibinfo {volume} {604}},\
  \bibinfo {pages} {451} (\bibinfo {year} {2022})}\BibitemShut {NoStop}%
\bibitem [{\citenamefont {Graham}\ \emph {et~al.}(2022)\citenamefont {Graham},
  \citenamefont {Song}, \citenamefont {Scott}, \citenamefont {Poole},
  \citenamefont {Phuttitarn}, \citenamefont {Jooya}, \citenamefont {Eichler},
  \citenamefont {Jiang}, \citenamefont {Marra}, \citenamefont {Grinkemeyer},
  \citenamefont {Kwon}, \citenamefont {Ebert}, \citenamefont {Cherek},
  \citenamefont {Lichtman}, \citenamefont {Gillette}, \citenamefont {Gilbert},
  \citenamefont {Bowman}, \citenamefont {Ballance}, \citenamefont {Campbell},
  \citenamefont {Dahl}, \citenamefont {Crawford}, \citenamefont {Blunt},
  \citenamefont {Rogers}, \citenamefont {Noel},\ and\ \citenamefont
  {Saffman}}]{Graham2022}%
  \BibitemOpen
  \bibfield  {author} {\bibinfo {author} {\bibfnamefont {T.~M.}\ \bibnamefont
  {Graham}}, \bibinfo {author} {\bibfnamefont {Y.}~\bibnamefont {Song}},
  \bibinfo {author} {\bibfnamefont {J.}~\bibnamefont {Scott}}, \bibinfo
  {author} {\bibfnamefont {C.}~\bibnamefont {Poole}}, \bibinfo {author}
  {\bibfnamefont {L.}~\bibnamefont {Phuttitarn}}, \bibinfo {author}
  {\bibfnamefont {K.}~\bibnamefont {Jooya}}, \bibinfo {author} {\bibfnamefont
  {P.}~\bibnamefont {Eichler}}, \bibinfo {author} {\bibfnamefont
  {X.}~\bibnamefont {Jiang}}, \bibinfo {author} {\bibfnamefont
  {A.}~\bibnamefont {Marra}}, \bibinfo {author} {\bibfnamefont
  {B.}~\bibnamefont {Grinkemeyer}}, \bibinfo {author} {\bibfnamefont
  {M.}~\bibnamefont {Kwon}}, \bibinfo {author} {\bibfnamefont {M.}~\bibnamefont
  {Ebert}}, \bibinfo {author} {\bibfnamefont {J.}~\bibnamefont {Cherek}},
  \bibinfo {author} {\bibfnamefont {M.~T.}\ \bibnamefont {Lichtman}}, \bibinfo
  {author} {\bibfnamefont {M.}~\bibnamefont {Gillette}}, \bibinfo {author}
  {\bibfnamefont {J.}~\bibnamefont {Gilbert}}, \bibinfo {author} {\bibfnamefont
  {D.}~\bibnamefont {Bowman}}, \bibinfo {author} {\bibfnamefont
  {T.}~\bibnamefont {Ballance}}, \bibinfo {author} {\bibfnamefont
  {C.}~\bibnamefont {Campbell}}, \bibinfo {author} {\bibfnamefont {E.~D.}\
  \bibnamefont {Dahl}}, \bibinfo {author} {\bibfnamefont {O.}~\bibnamefont
  {Crawford}}, \bibinfo {author} {\bibfnamefont {N.~S.}\ \bibnamefont {Blunt}},
  \bibinfo {author} {\bibfnamefont {B.}~\bibnamefont {Rogers}}, \bibinfo
  {author} {\bibfnamefont {T.}~\bibnamefont {Noel}}, \ and\ \bibinfo {author}
  {\bibfnamefont {M.}~\bibnamefont {Saffman}},\ }\bibfield  {title} {\enquote
  {\bibinfo {title} {{Multi-qubit entanglement and algorithms on a neutral-atom
  quantum computer}},}\ }\href {\doibase 10.1038/s41586-022-04603-6} {\bibfield
   {journal} {\bibinfo  {journal} {Nature}\ }\textbf {\bibinfo {volume}
  {604}},\ \bibinfo {pages} {457} (\bibinfo {year} {2022})}\BibitemShut
  {NoStop}%
\bibitem [{\citenamefont {Bloch}(2008)}]{Bloch2008}%
  \BibitemOpen
  \bibfield  {author} {\bibinfo {author} {\bibfnamefont {I.}~\bibnamefont
  {Bloch}},\ }\bibfield  {title} {\enquote {\bibinfo {title} {{Quantum
  coherence and entanglement with ultracold atoms in optical lattices}},}\
  }\href {\doibase 10.1038/nature07126} {\bibfield  {journal} {\bibinfo
  {journal} {Nature}\ }\textbf {\bibinfo {volume} {453}},\ \bibinfo {pages}
  {1016} (\bibinfo {year} {2008})}\BibitemShut {NoStop}%
\bibitem [{\citenamefont {Gross}\ and\ \citenamefont
  {Bloch}(2017)}]{Gross2017}%
  \BibitemOpen
  \bibfield  {author} {\bibinfo {author} {\bibfnamefont {C.}~\bibnamefont
  {Gross}}\ and\ \bibinfo {author} {\bibfnamefont {I.}~\bibnamefont {Bloch}},\
  }\bibfield  {title} {\enquote {\bibinfo {title} {{Quantum simulations with
  ultracold atoms in optical lattices}},}\ }\href {\doibase
  10.1126/science.aal3837} {\bibfield  {journal} {\bibinfo  {journal}
  {Science}\ }\textbf {\bibinfo {volume} {357}},\ \bibinfo {pages} {995}
  (\bibinfo {year} {2017})}\BibitemShut {NoStop}%
\bibitem [{\citenamefont {Yang}\ \emph
  {et~al.}(2020{\natexlab{a}})\citenamefont {Yang}, \citenamefont {Sun},
  \citenamefont {Ott}, \citenamefont {Wang}, \citenamefont {Zache},
  \citenamefont {Halimeh}, \citenamefont {Yuan}, \citenamefont {Hauke},\ and\
  \citenamefont {Pan}}]{Yang2020a}%
  \BibitemOpen
  \bibfield  {author} {\bibinfo {author} {\bibfnamefont {B.}~\bibnamefont
  {Yang}}, \bibinfo {author} {\bibfnamefont {H.}~\bibnamefont {Sun}}, \bibinfo
  {author} {\bibfnamefont {R.}~\bibnamefont {Ott}}, \bibinfo {author}
  {\bibfnamefont {H.-Y.}\ \bibnamefont {Wang}}, \bibinfo {author}
  {\bibfnamefont {T.~V.}\ \bibnamefont {Zache}}, \bibinfo {author}
  {\bibfnamefont {J.~C.}\ \bibnamefont {Halimeh}}, \bibinfo {author}
  {\bibfnamefont {Z.-S.}\ \bibnamefont {Yuan}}, \bibinfo {author}
  {\bibfnamefont {P.}~\bibnamefont {Hauke}}, \ and\ \bibinfo {author}
  {\bibfnamefont {J.-W.}\ \bibnamefont {Pan}},\ }\bibfield  {title} {\enquote
  {\bibinfo {title} {{Observation of gauge invariance in a 71-site
  Bose–Hubbard quantum simulator}},}\ }\href {\doibase
  10.1038/s41586-020-2910-8} {\bibfield  {journal} {\bibinfo  {journal}
  {Nature}\ }\textbf {\bibinfo {volume} {587}},\ \bibinfo {pages} {392}
  (\bibinfo {year} {2020}{\natexlab{a}})}\BibitemShut {NoStop}%
\bibitem [{\citenamefont {Yang}\ \emph
  {et~al.}(2020{\natexlab{b}})\citenamefont {Yang}, \citenamefont {Sun},
  \citenamefont {Huang}, \citenamefont {Wang}, \citenamefont {Deng},
  \citenamefont {Dai}, \citenamefont {Yuan},\ and\ \citenamefont
  {Pan}}]{Yang2020}%
  \BibitemOpen
  \bibfield  {author} {\bibinfo {author} {\bibfnamefont {B.}~\bibnamefont
  {Yang}}, \bibinfo {author} {\bibfnamefont {H.}~\bibnamefont {Sun}}, \bibinfo
  {author} {\bibfnamefont {C.-J.}\ \bibnamefont {Huang}}, \bibinfo {author}
  {\bibfnamefont {H.-Y.}\ \bibnamefont {Wang}}, \bibinfo {author}
  {\bibfnamefont {Y.}~\bibnamefont {Deng}}, \bibinfo {author} {\bibfnamefont
  {H.-N.}\ \bibnamefont {Dai}}, \bibinfo {author} {\bibfnamefont {Z.-S.}\
  \bibnamefont {Yuan}}, \ and\ \bibinfo {author} {\bibfnamefont {J.-W.}\
  \bibnamefont {Pan}},\ }\bibfield  {title} {\enquote {\bibinfo {title}
  {{Cooling and entangling ultracold atoms in optical lattices}},}\ }\href
  {\doibase 10.1126/science.aaz6801} {\bibfield  {journal} {\bibinfo  {journal}
  {Science}\ }\textbf {\bibinfo {volume} {369}},\ \bibinfo {pages} {550}
  (\bibinfo {year} {2020}{\natexlab{b}})}\BibitemShut {NoStop}%
\bibitem [{\citenamefont {Piotrowicz}\ \emph {et~al.}(2013)\citenamefont
  {Piotrowicz}, \citenamefont {Lichtman}, \citenamefont {Maller}, \citenamefont
  {Li}, \citenamefont {Zhang}, \citenamefont {Isenhower},\ and\ \citenamefont
  {Saffman}}]{Piotrowicz2013}%
  \BibitemOpen
  \bibfield  {author} {\bibinfo {author} {\bibfnamefont {M.~J.}\ \bibnamefont
  {Piotrowicz}}, \bibinfo {author} {\bibfnamefont {M.}~\bibnamefont
  {Lichtman}}, \bibinfo {author} {\bibfnamefont {K.}~\bibnamefont {Maller}},
  \bibinfo {author} {\bibfnamefont {G.}~\bibnamefont {Li}}, \bibinfo {author}
  {\bibfnamefont {S.}~\bibnamefont {Zhang}}, \bibinfo {author} {\bibfnamefont
  {L.}~\bibnamefont {Isenhower}}, \ and\ \bibinfo {author} {\bibfnamefont
  {M.}~\bibnamefont {Saffman}},\ }\bibfield  {title} {\enquote {\bibinfo
  {title} {{Two-dimensional lattice of blue-detuned atom traps using a
  projected Gaussian beam array}},}\ }\href {\doibase
  10.1103/PhysRevA.88.013420} {\bibfield  {journal} {\bibinfo  {journal}
  {Physical Review A}\ }\textbf {\bibinfo {volume} {88}},\ \bibinfo {pages}
  {013420} (\bibinfo {year} {2013})}\BibitemShut {NoStop}%
\bibitem [{\citenamefont {Campbell}\ \emph {et~al.}(2017)\citenamefont
  {Campbell}, \citenamefont {Hutson}, \citenamefont {Marti}, \citenamefont
  {Goban}, \citenamefont {{Darkwah Oppong}}, \citenamefont {McNally},
  \citenamefont {Sonderhouse}, \citenamefont {Robinson}, \citenamefont {Zhang},
  \citenamefont {Bloom},\ and\ \citenamefont {Ye}}]{Campbell2017}%
  \BibitemOpen
  \bibfield  {author} {\bibinfo {author} {\bibfnamefont {S.~L.}\ \bibnamefont
  {Campbell}}, \bibinfo {author} {\bibfnamefont {R.~B.}\ \bibnamefont
  {Hutson}}, \bibinfo {author} {\bibfnamefont {G.~E.}\ \bibnamefont {Marti}},
  \bibinfo {author} {\bibfnamefont {A.}~\bibnamefont {Goban}}, \bibinfo
  {author} {\bibfnamefont {N.}~\bibnamefont {{Darkwah Oppong}}}, \bibinfo
  {author} {\bibfnamefont {R.~L.}\ \bibnamefont {McNally}}, \bibinfo {author}
  {\bibfnamefont {L.}~\bibnamefont {Sonderhouse}}, \bibinfo {author}
  {\bibfnamefont {J.~M.}\ \bibnamefont {Robinson}}, \bibinfo {author}
  {\bibfnamefont {W.}~\bibnamefont {Zhang}}, \bibinfo {author} {\bibfnamefont
  {B.~J.}\ \bibnamefont {Bloom}}, \ and\ \bibinfo {author} {\bibfnamefont
  {J.}~\bibnamefont {Ye}},\ }\bibfield  {title} {\enquote {\bibinfo {title} {{A
  Fermi-degenerate three-dimensional optical lattice clock}},}\ }\href
  {\doibase 10.1126/science.aam5538} {\bibfield  {journal} {\bibinfo  {journal}
  {Science}\ }\textbf {\bibinfo {volume} {358}},\ \bibinfo {pages} {90}
  (\bibinfo {year} {2017})}\BibitemShut {NoStop}%
\bibitem [{\citenamefont {Nemitz}\ \emph {et~al.}(2016)\citenamefont {Nemitz},
  \citenamefont {Ohkubo}, \citenamefont {Takamoto}, \citenamefont {Ushijima},
  \citenamefont {Das}, \citenamefont {Ohmae},\ and\ \citenamefont
  {Katori}}]{Nemitz2016}%
  \BibitemOpen
  \bibfield  {author} {\bibinfo {author} {\bibfnamefont {N.}~\bibnamefont
  {Nemitz}}, \bibinfo {author} {\bibfnamefont {T.}~\bibnamefont {Ohkubo}},
  \bibinfo {author} {\bibfnamefont {M.}~\bibnamefont {Takamoto}}, \bibinfo
  {author} {\bibfnamefont {I.}~\bibnamefont {Ushijima}}, \bibinfo {author}
  {\bibfnamefont {M.}~\bibnamefont {Das}}, \bibinfo {author} {\bibfnamefont
  {N.}~\bibnamefont {Ohmae}}, \ and\ \bibinfo {author} {\bibfnamefont
  {H.}~\bibnamefont {Katori}},\ }\bibfield  {title} {\enquote {\bibinfo {title}
  {{Frequency ratio of Yb and Sr clocks with ${5}\,\times\,{10^{-17}}$
  uncertainty at 150 seconds averaging time}},}\ }\href {\doibase
  10.1038/nphoton.2016.20} {\bibfield  {journal} {\bibinfo  {journal} {Nature
  Photonics}\ }\textbf {\bibinfo {volume} {10}},\ \bibinfo {pages} {258}
  (\bibinfo {year} {2016})}\BibitemShut {NoStop}%
\bibitem [{\citenamefont {Takano}\ \emph {et~al.}(2016)\citenamefont {Takano},
  \citenamefont {Takamoto}, \citenamefont {Ushijima}, \citenamefont {Ohmae},
  \citenamefont {Akatsuka}, \citenamefont {Yamaguchi}, \citenamefont
  {Kuroishi}, \citenamefont {Munekane}, \citenamefont {Miyahara},\ and\
  \citenamefont {Katori}}]{Takano2016}%
  \BibitemOpen
  \bibfield  {author} {\bibinfo {author} {\bibfnamefont {T.}~\bibnamefont
  {Takano}}, \bibinfo {author} {\bibfnamefont {M.}~\bibnamefont {Takamoto}},
  \bibinfo {author} {\bibfnamefont {I.}~\bibnamefont {Ushijima}}, \bibinfo
  {author} {\bibfnamefont {N.}~\bibnamefont {Ohmae}}, \bibinfo {author}
  {\bibfnamefont {T.}~\bibnamefont {Akatsuka}}, \bibinfo {author}
  {\bibfnamefont {A.}~\bibnamefont {Yamaguchi}}, \bibinfo {author}
  {\bibfnamefont {Y.}~\bibnamefont {Kuroishi}}, \bibinfo {author}
  {\bibfnamefont {H.}~\bibnamefont {Munekane}}, \bibinfo {author}
  {\bibfnamefont {B.}~\bibnamefont {Miyahara}}, \ and\ \bibinfo {author}
  {\bibfnamefont {H.}~\bibnamefont {Katori}},\ }\bibfield  {title} {\enquote
  {\bibinfo {title} {{Geopotential measurements with synchronously linked
  optical lattice clocks}},}\ }\href {\doibase 10.1038/nphoton.2016.159}
  {\bibfield  {journal} {\bibinfo  {journal} {Nature Photonics}\ }\textbf
  {\bibinfo {volume} {10}},\ \bibinfo {pages} {662} (\bibinfo {year}
  {2016})}\BibitemShut {NoStop}%
\bibitem [{\citenamefont {Wang}\ \emph {et~al.}(2023)\citenamefont {Wang},
  \citenamefont {Gu}, \citenamefont {Hu}, \citenamefont {Zhang}, \citenamefont
  {Zhang}, \citenamefont {Li}, \citenamefont {He}, \citenamefont {Zou},
  \citenamefont {Dong}, \citenamefont {Guo},\ and\ \citenamefont
  {Zou}}]{Wang2023}%
  \BibitemOpen
  \bibfield  {author} {\bibinfo {author} {\bibfnamefont {Z.-B.}\ \bibnamefont
  {Wang}}, \bibinfo {author} {\bibfnamefont {C.}~\bibnamefont {Gu}}, \bibinfo
  {author} {\bibfnamefont {X.-X.}\ \bibnamefont {Hu}}, \bibinfo {author}
  {\bibfnamefont {Y.-T.}\ \bibnamefont {Zhang}}, \bibinfo {author}
  {\bibfnamefont {J.-Z.}\ \bibnamefont {Zhang}}, \bibinfo {author}
  {\bibfnamefont {G.}~\bibnamefont {Li}}, \bibinfo {author} {\bibfnamefont
  {X.}~\bibnamefont {He}}, \bibinfo {author} {\bibfnamefont {X.-B.}\
  \bibnamefont {Zou}}, \bibinfo {author} {\bibfnamefont {C.}~\bibnamefont
  {Dong}}, \bibinfo {author} {\bibfnamefont {G.-c.}\ \bibnamefont {Guo}}, \
  and\ \bibinfo {author} {\bibfnamefont {C.-L.}\ \bibnamefont {Zou}},\
  }\bibfield  {title} {\enquote {\bibinfo {title} {{Controllable atomic
  collision in a tight optical dipole trap}},}\ }\href {\doibase
  10.1364/OL.479036} {\bibfield  {journal} {\bibinfo  {journal} {Optics
  Letters}\ }\textbf {\bibinfo {volume} {48}},\ \bibinfo {pages} {1064}
  (\bibinfo {year} {2023})}\BibitemShut {NoStop}%
\bibitem [{\citenamefont {Vochezer}\ \emph {et~al.}(2018)\citenamefont
  {Vochezer}, \citenamefont {Kampschulte}, \citenamefont {Hammerer},\ and\
  \citenamefont {Treutlein}}]{Vochezer2018}%
  \BibitemOpen
  \bibfield  {author} {\bibinfo {author} {\bibfnamefont {A.}~\bibnamefont
  {Vochezer}}, \bibinfo {author} {\bibfnamefont {T.}~\bibnamefont
  {Kampschulte}}, \bibinfo {author} {\bibfnamefont {K.}~\bibnamefont
  {Hammerer}}, \ and\ \bibinfo {author} {\bibfnamefont {P.}~\bibnamefont
  {Treutlein}},\ }\bibfield  {title} {\enquote {\bibinfo {title}
  {{Light-Mediated Collective Atomic Motion in an Optical Lattice Coupled to a
  Membrane}},}\ }\href {\doibase 10.1103/PhysRevLett.120.073602} {\bibfield
  {journal} {\bibinfo  {journal} {Physical Review Letters}\ }\textbf {\bibinfo
  {volume} {120}},\ \bibinfo {pages} {073602} (\bibinfo {year}
  {2018})}\BibitemShut {NoStop}%
\bibitem [{\citenamefont {Chen}\ \emph {et~al.}(2023)\citenamefont {Chen},
  \citenamefont {Zhao}, \citenamefont {Wang}, \citenamefont {Li}, \citenamefont
  {Zhang}, \citenamefont {Chen}, \citenamefont {Zhang}, \citenamefont {Xu},
  \citenamefont {Liu}, \citenamefont {Dong}, \citenamefont {Guo}, \citenamefont
  {Huang},\ and\ \citenamefont {Zou}}]{ChenGJ2023}%
  \BibitemOpen
  \bibfield  {author} {\bibinfo {author} {\bibfnamefont {G.-J.}\ \bibnamefont
  {Chen}}, \bibinfo {author} {\bibfnamefont {D.}~\bibnamefont {Zhao}}, \bibinfo
  {author} {\bibfnamefont {Z.-B.}\ \bibnamefont {Wang}}, \bibinfo {author}
  {\bibfnamefont {Z.}~\bibnamefont {Li}}, \bibinfo {author} {\bibfnamefont
  {J.-Z.}\ \bibnamefont {Zhang}}, \bibinfo {author} {\bibfnamefont
  {L.}~\bibnamefont {Chen}}, \bibinfo {author} {\bibfnamefont {Y.-L.}\
  \bibnamefont {Zhang}}, \bibinfo {author} {\bibfnamefont {X.-B.}\ \bibnamefont
  {Xu}}, \bibinfo {author} {\bibfnamefont {A.-P.}\ \bibnamefont {Liu}},
  \bibinfo {author} {\bibfnamefont {C.-H.}\ \bibnamefont {Dong}}, \bibinfo
  {author} {\bibfnamefont {G.-C.}\ \bibnamefont {Guo}}, \bibinfo {author}
  {\bibfnamefont {K.}~\bibnamefont {Huang}}, \ and\ \bibinfo {author}
  {\bibfnamefont {C.-L.}\ \bibnamefont {Zou}},\ }\bibfield  {title} {\enquote
  {\bibinfo {title} {{Trapping and characterizing single atom using
  metalens}},}\ }\href@noop {} {\bibfield  {journal} {\bibinfo  {journal} {In
  preparation}\ } (\bibinfo {year} {2023})}\BibitemShut {NoStop}%
\bibitem [{\citenamefont {Hsu}\ \emph {et~al.}(2022)\citenamefont {Hsu},
  \citenamefont {Zhu}, \citenamefont {Thiele}, \citenamefont {Brown},
  \citenamefont {Papp}, \citenamefont {Agrawal},\ and\ \citenamefont
  {Regal}}]{Hsu2022}%
  \BibitemOpen
  \bibfield  {author} {\bibinfo {author} {\bibfnamefont {T.-W.}\ \bibnamefont
  {Hsu}}, \bibinfo {author} {\bibfnamefont {W.}~\bibnamefont {Zhu}}, \bibinfo
  {author} {\bibfnamefont {T.}~\bibnamefont {Thiele}}, \bibinfo {author}
  {\bibfnamefont {M.~O.}\ \bibnamefont {Brown}}, \bibinfo {author}
  {\bibfnamefont {S.~B.}\ \bibnamefont {Papp}}, \bibinfo {author}
  {\bibfnamefont {A.}~\bibnamefont {Agrawal}}, \ and\ \bibinfo {author}
  {\bibfnamefont {C.~A.}\ \bibnamefont {Regal}},\ }\bibfield  {title} {\enquote
  {\bibinfo {title} {{Single-Atom Trapping in a Metasurface-Lens Optical
  Tweezer}},}\ }\href {\doibase 10.1103/PRXQuantum.3.030316} {\bibfield
  {journal} {\bibinfo  {journal} {PRX Quantum}\ }\textbf {\bibinfo {volume}
  {3}},\ \bibinfo {pages} {030316} (\bibinfo {year} {2022})}\BibitemShut
  {NoStop}%
\bibitem [{\citenamefont {Kim}\ \emph {et~al.}(2019)\citenamefont {Kim},
  \citenamefont {Chang}, \citenamefont {Fields}, \citenamefont {Chen},\ and\
  \citenamefont {Hung}}]{Kim}%
  \BibitemOpen
  \bibfield  {author} {\bibinfo {author} {\bibfnamefont {M.~E.}\ \bibnamefont
  {Kim}}, \bibinfo {author} {\bibfnamefont {T.-H.}\ \bibnamefont {Chang}},
  \bibinfo {author} {\bibfnamefont {B.~M.}\ \bibnamefont {Fields}}, \bibinfo
  {author} {\bibfnamefont {C.-A.}\ \bibnamefont {Chen}}, \ and\ \bibinfo
  {author} {\bibfnamefont {C.-L.}\ \bibnamefont {Hung}},\ }\bibfield  {title}
  {\enquote {\bibinfo {title} {{Trapping single atoms on a nanophotonic circuit
  with configurable tweezer lattices}},}\ }\href {\doibase
  10.1038/s41467-019-09635-7} {\bibfield  {journal} {\bibinfo  {journal}
  {Nature Communications}\ }\textbf {\bibinfo {volume} {10}},\ \bibinfo {pages}
  {1647} (\bibinfo {year} {2019})}\BibitemShut {NoStop}%
\bibitem [{\citenamefont {Schymik}\ \emph {et~al.}(2022)\citenamefont
  {Schymik}, \citenamefont {Ximenez}, \citenamefont {Bloch}, \citenamefont
  {Dreon}, \citenamefont {Signoles}, \citenamefont {Nogrette}, \citenamefont
  {Barredo}, \citenamefont {Browaeys},\ and\ \citenamefont
  {Lahaye}}]{Schymik2022}%
  \BibitemOpen
  \bibfield  {author} {\bibinfo {author} {\bibfnamefont {K.~N.}\ \bibnamefont
  {Schymik}}, \bibinfo {author} {\bibfnamefont {B.}~\bibnamefont {Ximenez}},
  \bibinfo {author} {\bibfnamefont {E.}~\bibnamefont {Bloch}}, \bibinfo
  {author} {\bibfnamefont {D.}~\bibnamefont {Dreon}}, \bibinfo {author}
  {\bibfnamefont {A.}~\bibnamefont {Signoles}}, \bibinfo {author}
  {\bibfnamefont {F.}~\bibnamefont {Nogrette}}, \bibinfo {author}
  {\bibfnamefont {D.}~\bibnamefont {Barredo}}, \bibinfo {author} {\bibfnamefont
  {A.}~\bibnamefont {Browaeys}}, \ and\ \bibinfo {author} {\bibfnamefont
  {T.}~\bibnamefont {Lahaye}},\ }\bibfield  {title} {\enquote {\bibinfo {title}
  {{In situ equalization of single-atom loading in large-scale optical tweezer
  arrays}},}\ }\href {\doibase 10.1103/PhysRevA.106.022611} {\bibfield
  {journal} {\bibinfo  {journal} {Physical Review A}\ }\textbf {\bibinfo
  {volume} {106}},\ \bibinfo {pages} {022611} (\bibinfo {year}
  {2022})}\BibitemShut {NoStop}%
\bibitem [{\citenamefont {Hilliard}\ \emph {et~al.}(2015)\citenamefont
  {Hilliard}, \citenamefont {Fung}, \citenamefont {Sompet}, \citenamefont
  {Carpentier},\ and\ \citenamefont {Andersen}}]{Hilliard2015}%
  \BibitemOpen
  \bibfield  {author} {\bibinfo {author} {\bibfnamefont {A.~J.}\ \bibnamefont
  {Hilliard}}, \bibinfo {author} {\bibfnamefont {Y.~H.}\ \bibnamefont {Fung}},
  \bibinfo {author} {\bibfnamefont {P.}~\bibnamefont {Sompet}}, \bibinfo
  {author} {\bibfnamefont {A.~V.}\ \bibnamefont {Carpentier}}, \ and\ \bibinfo
  {author} {\bibfnamefont {M.~F.}\ \bibnamefont {Andersen}},\ }\bibfield
  {title} {\enquote {\bibinfo {title} {{In-trap fluorescence detection of atoms
  in a microscopic dipole trap}},}\ }\href {\doibase
  10.1103/PhysRevA.91.053414} {\bibfield  {journal} {\bibinfo  {journal}
  {Physical Review A}\ }\textbf {\bibinfo {volume} {91}},\ \bibinfo {pages}
  {053414} (\bibinfo {year} {2015})}\BibitemShut {NoStop}%
\bibitem [{\citenamefont {Hu}\ \emph {et~al.}(2019)\citenamefont {Hu},
  \citenamefont {Zhao}, \citenamefont {Wang}, \citenamefont {Zhang},
  \citenamefont {Zou}, \citenamefont {Dong}, \citenamefont {Tang},
  \citenamefont {Guo},\ and\ \citenamefont {Zou}}]{Hu2019}%
  \BibitemOpen
  \bibfield  {author} {\bibinfo {author} {\bibfnamefont {X.-X.}\ \bibnamefont
  {Hu}}, \bibinfo {author} {\bibfnamefont {C.-L.}\ \bibnamefont {Zhao}},
  \bibinfo {author} {\bibfnamefont {Z.-B.}\ \bibnamefont {Wang}}, \bibinfo
  {author} {\bibfnamefont {Y.-L.}\ \bibnamefont {Zhang}}, \bibinfo {author}
  {\bibfnamefont {X.-B.}\ \bibnamefont {Zou}}, \bibinfo {author} {\bibfnamefont
  {C.-H.}\ \bibnamefont {Dong}}, \bibinfo {author} {\bibfnamefont {H.~X.}\
  \bibnamefont {Tang}}, \bibinfo {author} {\bibfnamefont {G.-C.}\ \bibnamefont
  {Guo}}, \ and\ \bibinfo {author} {\bibfnamefont {C.-L.}\ \bibnamefont
  {Zou}},\ }\bibfield  {title} {\enquote {\bibinfo {title} {Cavity-enhanced
  optical controlling based on three-wave mixing in cavity-atom ensemble
  system},}\ }\href {\doibase 10.1364/oe.27.006660} {\bibfield  {journal}
  {\bibinfo  {journal} {Optics Express}\ }\textbf {\bibinfo {volume} {27}},\
  \bibinfo {pages} {6660} (\bibinfo {year} {2019})}\BibitemShut {NoStop}%
\bibitem [{\citenamefont {Bothwell}\ \emph {et~al.}(2022)\citenamefont
  {Bothwell}, \citenamefont {Kennedy}, \citenamefont {Aeppli}, \citenamefont
  {Kedar}, \citenamefont {Robinson}, \citenamefont {Oelker}, \citenamefont
  {Staron},\ and\ \citenamefont {Ye}}]{Bothwell2022}%
  \BibitemOpen
  \bibfield  {author} {\bibinfo {author} {\bibfnamefont {T.}~\bibnamefont
  {Bothwell}}, \bibinfo {author} {\bibfnamefont {C.~J.}\ \bibnamefont
  {Kennedy}}, \bibinfo {author} {\bibfnamefont {A.}~\bibnamefont {Aeppli}},
  \bibinfo {author} {\bibfnamefont {D.}~\bibnamefont {Kedar}}, \bibinfo
  {author} {\bibfnamefont {J.~M.}\ \bibnamefont {Robinson}}, \bibinfo {author}
  {\bibfnamefont {E.}~\bibnamefont {Oelker}}, \bibinfo {author} {\bibfnamefont
  {A.}~\bibnamefont {Staron}}, \ and\ \bibinfo {author} {\bibfnamefont
  {J.}~\bibnamefont {Ye}},\ }\bibfield  {title} {\enquote {\bibinfo {title}
  {{Resolving the gravitational redshift across a millimetre-scale atomic
  sample}},}\ }\href {\doibase 10.1038/s41586-021-04349-7} {\bibfield
  {journal} {\bibinfo  {journal} {Nature}\ }\textbf {\bibinfo {volume} {602}},\
  \bibinfo {pages} {420} (\bibinfo {year} {2022})}\BibitemShut {NoStop}%
\bibitem [{\citenamefont {Young}\ \emph {et~al.}(2020)\citenamefont {Young},
  \citenamefont {Eckner}, \citenamefont {Milner}, \citenamefont {Kedar},
  \citenamefont {Norcia}, \citenamefont {Oelker}, \citenamefont {Schine},
  \citenamefont {Ye},\ and\ \citenamefont {Kaufman}}]{Young2020}%
  \BibitemOpen
  \bibfield  {author} {\bibinfo {author} {\bibfnamefont {A.~W.}\ \bibnamefont
  {Young}}, \bibinfo {author} {\bibfnamefont {W.~J.}\ \bibnamefont {Eckner}},
  \bibinfo {author} {\bibfnamefont {W.~R.}\ \bibnamefont {Milner}}, \bibinfo
  {author} {\bibfnamefont {D.}~\bibnamefont {Kedar}}, \bibinfo {author}
  {\bibfnamefont {M.~A.}\ \bibnamefont {Norcia}}, \bibinfo {author}
  {\bibfnamefont {E.}~\bibnamefont {Oelker}}, \bibinfo {author} {\bibfnamefont
  {N.}~\bibnamefont {Schine}}, \bibinfo {author} {\bibfnamefont
  {J.}~\bibnamefont {Ye}}, \ and\ \bibinfo {author} {\bibfnamefont {A.~M.}\
  \bibnamefont {Kaufman}},\ }\bibfield  {title} {\enquote {\bibinfo {title}
  {{Half-minute-scale atomic coherence and high relative stability in a tweezer
  clock}},}\ }\href {\doibase 10.1038/s41586-020-3009-y} {\bibfield  {journal}
  {\bibinfo  {journal} {Nature}\ }\textbf {\bibinfo {volume} {588}},\ \bibinfo
  {pages} {408} (\bibinfo {year} {2020})}\BibitemShut {NoStop}%
\bibitem [{\citenamefont {Norcia}\ \emph {et~al.}(2019)\citenamefont {Norcia},
  \citenamefont {Young}, \citenamefont {Eckner}, \citenamefont {Oelker},
  \citenamefont {Ye},\ and\ \citenamefont {Kaufman}}]{Norcia2019}%
  \BibitemOpen
  \bibfield  {author} {\bibinfo {author} {\bibfnamefont {M.~A.}\ \bibnamefont
  {Norcia}}, \bibinfo {author} {\bibfnamefont {A.~W.}\ \bibnamefont {Young}},
  \bibinfo {author} {\bibfnamefont {W.~J.}\ \bibnamefont {Eckner}}, \bibinfo
  {author} {\bibfnamefont {E.}~\bibnamefont {Oelker}}, \bibinfo {author}
  {\bibfnamefont {J.}~\bibnamefont {Ye}}, \ and\ \bibinfo {author}
  {\bibfnamefont {A.~M.}\ \bibnamefont {Kaufman}},\ }\bibfield  {title}
  {\enquote {\bibinfo {title} {{Seconds-scale coherence on an optical clock
  transition in a tweezer array}},}\ }\href {\doibase 10.1126/science.aay0644}
  {\bibfield  {journal} {\bibinfo  {journal} {Science}\ }\textbf {\bibinfo
  {volume} {366}},\ \bibinfo {pages} {93} (\bibinfo {year} {2019})}\BibitemShut
  {NoStop}%
\end{thebibliography}%

\end{document}